\documentclass[12pt,twoside,fleqn]{article}
\usepackage{color,colortbl,graphicx,setspace,multirow,enumitem}

\usepackage[utf8, latin1]{inputenc}

\usepackage{geometry}
\usepackage[hang,flushmargin]{footmisc} 
\usepackage[hang]{footmisc}
\usepackage{natbib} 
\usepackage{bibunits}
\defaultbibliography{references}
\defaultbibliographystyle{custom}

\usepackage{amsfonts}
\usepackage{amsmath}
\usepackage{amssymb}
\usepackage{placeins}

\let\emptyset\varnothing

\usepackage{booktabs,threeparttable} 

\usepackage[dvipsnames]{xcolor}
\definecolor{nblue}{HTML}{000660}
\usepackage[colorlinks=true,urlcolor=nblue,linkcolor=nblue,citecolor=nblue]{hyperref}
\newcommand{\inlinehead}[1]{\textsc{\sffamily\textbf{#1}}. }

\newcommand{\diag}{\text{diag}}

\newcommand{\tts}{\texttt{s}}
\newcommand{\ttb}{\texttt{b}}

\usepackage[title,titletoc]{appendix}

\usepackage{titlesec} 
\titleformat{\section}[block]{\bfseries\sffamily\large}{\thesection. }{0em}{\MakeUppercase} 
\titleformat{\subsection}[block]{\bfseries\sffamily\large}{\thesubsection. }{0em}{} 
\titleformat{\subsubsection}[block]{\large\sffamily}{}{0em}{\itshape} 



\usepackage{bm}
\usepackage{caption}
\usepackage{subcaption}
\newcommand\cond[1][]{\:#1\vert\:}

\makeatletter\let\p@subfigure\thefigure\makeatother

\usepackage{fancyhdr} 
\def\titletext{Scenario Analysis with Multivariate Bayesian Machine Learning Models}

\title{\sffamily\huge{\textbf{\titletext}}}
\author{}
\date{}

\begin{document}

\maketitle
\vspace*{-4.5em} 
\normalsize
\begin{center}
\begin{minipage}{.49\textwidth}
  \large\centering Michael \MakeUppercase{Pfarrhofer}\\[0.25em]
  \small WU Vienna
\end{minipage}
\begin{minipage}{.49\textwidth}
  \large\centering Anna \MakeUppercase{Stelzer}\\[0.25em]
  \small Oesterreichische Nationalbank
\end{minipage}
\end{center}

\vspace*{1em}
\doublespacing
\begin{center}
\begin{minipage}{0.8\textwidth}
\noindent\small We present an econometric framework that adapts tools for scenario analysis, such as variants of conditional forecasts and generalized impulse responses, for use with dynamic nonparametric models. The proposed algorithms are based on predictive simulation and sequential Monte Carlo methods. Their utility is demonstrated with three applications: (1) conditional forecasts based on stress test scenarios, measuring (2) macroeconomic risk under varying financial stress, and estimating the (3) asymmetric effects of financial shocks in the US and their international spillovers. Our empirical results indicate the importance of nonlinearities and asymmetries in relationships between macroeconomic and financial variables.

\end{minipage}
\end{center}

\singlespacing\vfill\noindent{\footnotesize\textit{Contact}: \href{mailto:michael.pfarrhofer@wu.ac.at}{michael.pfarrhofer@wu.ac.at}, Department of Economics, WU Vienna University of Economics and Business. We thank Christiane Baumeister, Niko Hauzenberger, Philip Lane, Massimiliano Marcellino, and seminar participants at the OeNB, Swiss National Bank (SNB), and IMIM seminar for useful comments. \textit{Disclaimer:} The views expressed in this paper do not necessarily reflect those of the Oesterreichische Nationalbank or the Eurosystem.}

\thispagestyle{empty}\renewcommand{\footnotelayout}{\setstretch{1.5}}\newpage
\doublespacing\normalsize\renewcommand{\thepage}{\arabic{page}}

\begin{bibunit}
\section{Introduction}\label{sec:introduction}
In this paper, we discuss how to conduct scenario analysis with dynamic multivariate models in macroeconomics when the functional form of the conditional mean is nonlinear and/or unknown. While related tools exist in linear and traditional nonlinear frameworks (e.g., regime switching or time-varying parameter models, see \citet{fischer2023general} for a recent example), they rely on potentially restrictive parametric assumptions and are not directly applicable in nonparametric models. The latter feature in many recent macroeconometric applications that employ Bayesian machine learning (ML) methods \citep[see, e.g.,][]{huber2022inference,clark2021tail,huber2023nowcasting,hauzenberger2024gaussian,chernis2025bayesian,lima2025minnesota}.

We use the term scenario analysis broadly to refer to different counterfactual experiments. These include versions of conditional forecasts (CFs) and structural scenario analysis \citep[as in][]{antolin2021structural}, and several variants of nonlinear impulse response functions (IRFs). While some aspects and issues in the Bayesian ML context have been discussed in isolation in the aforementioned papers, a unified or comprehensive treatment is not yet available. We bridge this gap by developing a framework and estimation algorithms for scenario analysis that are widely applicable when the conditional mean is flexibly learned from the data.

Conditional forecasts simulate future paths of variables under scenarios encoded as constraints on other observables, structural shocks, or both. Related techniques have been used by academics and practitioners since the 1980s in linear models \citep[see, e.g.,][]{doan1984forecasting}, with subsequent refinements \citep[see][]{waggoner1999conditional,andersson2010density,jarocinski2010conditional,banbura2015conditional,antolin2021structural,chernis2024decision,chan2023conditional}.\footnote{For a Bayesian decision analysis perspective and discussion of additional related literature, see \citet{west2024perspectives}.} Breaking the assumption of linearity, however, complicates matters: there is no general representation of nonlinear time series as functions of shocks \citep[see, e.g.,][]{potter2000nonlinear}, and unknown nonlinearities make it difficult to derive multi-step ahead predictive distributions. Obtaining closed form solutions is often impossible, but in this case one can resort to Monte Carlo methods. We show how common approaches can be adapted to nonparametric settings. Akin to \citet{banbura2015conditional}, our approach casts conditional forecasting as a nonlinear state space problem. We use Particle Gibbs with Ancestor Sampling \citep[PGAS,][]{lindsten14a}, which combines generality with computational efficiency, making it suitable for a comparatively wide range of nonlinear multivariate models.

An IRF can be defined as the difference between conditional forecasts \citep{gallant1993nonlinear,koop1996impulse,rambachan2021common,jordalocal}. Our framework thus lends itself to estimating nonlinear dynamic effects \citep[see also][for related discussions]{gonccalves2021impulse,gonccalves2024state,kolesar2024dynamic} in the form of generalized IRFs (GIRFs), and thereby contributes to improving interpretability of ML methods. Compared with IRFs in linear settings (which are symmetric, shape invariant and history independent) the GIRFs do not feature such potentially restrictive properties. Specifically, we consider (1) unorthogonalized GIRFs (the difference between conditional and unconditional forecasts subject to certain restrictions on observables), we explore (2) structural GIRFs to shocks identified with approaches typically used in the structural VAR (SVAR) literature, and (3) restricted GIRFs, which can be used to construct policy counterfactuals and quantify the contributions of specific sets of transmission channels in the propagation of shocks.\footnote{A related approach derived from \citet{antolin2021structural} is discussed in \citet{breitenlechner2022goes}. Counterfactual experiments in this spirit relate to alternative policy rules and are subject to the \citet{lucas1976} critique \citep[for a recent discussion, see][]{mckay2023can}. While our baseline approach is also susceptible to these concerns, there are options to improve robustness, and nonparametric modeling offers additional flexibility and a safeguard; see Section \ref{sec:GIRFs} for a more detailed discussion.}

The proposed approach is generally applicable in multivariate dynamic models. This means that there are many potential candidates for estimating conditional mean functions. These choices include (but are not limited to) regression trees or Gaussian process priors \citep[see][for a recent review]{marcellino2024bookchapter}. Such nonparametric methods are more robust to misspecification that may arise from assuming tightly parameterized likelihoods, and can be used to capturing genuinely nonlinear patterns in the data. For our empirical work, we use Bayesian Additive Regression Trees \citep[BART,][]{chipman2010bart} as a specific nonparametric implementation. We pick this sum-of-trees model because tree-based approaches have proven particularly capable of producing accurate forecasts when used with time series for the US economy, with datasets structured similarly to the one we use in this paper \citep[see, e.g.,][]{medeiros2021forecasting,goulet2022machine,clark2021tail,goulet2024bag}. Our approach to predictive inference is developed under the assumption of an additive multivariate Gaussian error term with a time-varying variance covariance matrix.

We illustrate our framework through several empirical applications. Our dataset comprises about $25$ quarterly macroeconomic and financial variables for the US economy ranging from the mid-$1970$s to the last quarter of $2024$. In one of our explorations, we also add some international variables for the euro area (EA) and the United Kingdom (UK). The applied work assesses and illustrates the role of nonlinearities when interest centers on CFs, and we explore asymmetries in the propagation of shocks of different types, signs and magnitudes. Specifically, we provide three empirical applications. First, inspired by \citet{chan2023conditional}, we use a subset of the macroeconomic assumptions underlying the annual stress test conducted by the Federal Reserve System and compute CFs using constraints on observables for different scenarios, comparing predictive densities from linear and nonlinear models. Second, reflecting the growth-at-risk literature \citep{adrian2019vulnerable}, we study the unorthogonalized counterfactual implications of varying financial conditions on tail risks of output growth, inflation, and employment. Third, we identify a US-based financial shock recursively \citep[as in, e.g.,][]{barnichon2022effects}, and compute GIRFs to shocks of different signs and magnitudes that are allowed to propagate internationally. We then use a restricted GIRF approach to gauge the role of spillovers and spillbacks, inspired by the empirical work of \citet{breitenlechner2022goes}.

The rest of this paper is structured as follows. Section \ref{sec:predictions} addresses the challenges and solutions to obtain predictive inference in the presence of nonlinearities of unknown form. We discuss how to impose constraints on forecasts, and how these constraints may be used to construct scenarios through GIRFs. Section \ref{sec:econometrics} presents an econometric implementation using BART. Section \ref{sec:application} provides three empirical illustrations. The last section concludes.

\section{Nonlinearity and Predictive distributions}\label{sec:predictions}
Let $\bm{y}_t = (y_{1t},\hdots,y_{nt})'$ collect $n$ variables for $t = 1,\hdots, T,$ and $\bm{x}_t = (\bm{y}_{t-1}',\hdots,\bm{y}_{t-p}')'$ is a $k=np$ vector of lags. Interest centers on dynamic multivariate models of the form:
\begin{equation}
\bm{y}_t = \bm{F}(\bm{x}_t) + \bm{\epsilon}_t, \quad \bm{\epsilon}_t \sim \mathcal{N}(\bm{0}_n,\bm{\Sigma}_t),\label{eq:general-model}
\end{equation}
where $\bm{F}(\bm{x}_t) = (f_1(\bm{x}_t),\hdots,f_n(\bm{x}_t))'$ is an $n$-vector of conditional mean functions $f_i(\bm{x}_t):\mathbb{R}^k\rightarrow\mathbb{R}$ for $i=1,\hdots,n,$ such that $\bm{F}(\bm{x}_t):\mathbb{R}^k\rightarrow\mathbb{R}^n$. We assume iid reduced form Gaussian errors $\bm{\epsilon}_t$, with $n\times n$ time-varying covariance matrix $\bm{\Sigma}_t$. One may assume functional forms for the $f_i(\bm{x}_t)$'s or treat them as unknown and estimate them nonparametrically. The methods we propose are designed specifically for the latter case. 

All scenario analyses we discuss in this paper rely on computing functions of the (conditional) moments of $\bm{y}_{t+h}$ based on (\ref{eq:general-model}). 
Consider the single-lag case $p = 1$, without loss of generality. We may write:
\begin{equation}
    \bm{y}_{t+h} = \tilde{\bm{F}}_{(h)}(\bm{y}_t,\bm{\epsilon}_{t+1},\bm{\epsilon}_{t+2},\hdots,\bm{\epsilon}_{t+h}) = \bm{F}(\tilde{\bm{F}}_{(h-1)}(\bm{y}_t,\bm{\epsilon}_{t+1},\hdots,\bm{\epsilon}_{t+(h-1)})) + \bm{\epsilon}_{t+h},\label{eq:Fhpred}
\end{equation}
with $\bm{y}_{t+1} = \tilde{\bm{F}}_{(1)}(\bm{y}_t,\bm{\epsilon}_{t+1}) = \bm{F}(\bm{y}_{t}) + \bm{\epsilon}_{t+1}$, and $\tilde{\bm{F}}_{(h)}(\bullet)$ denotes the $h$-step composition of $\bm{F}(\bullet)$ which is defined recursively. This expression is related to the Wold decomposition, and expresses $\bm{y}_{t+h}$ as a function of the initial condition $\bm{y}_t$ and a sequence of white noise shocks, $\{\bm{\epsilon}_{t+1},\hdots,\bm{\epsilon}_{t+h}\}$; see also \citet{gourieroux2023nonlinear} and Appendix \ref{app:technical}.\footnote{Under some assumptions about the behavior of $\tilde{\bm{F}}_{(h)}(\bullet)$, it is noteworthy that expansions such as the Volterra series could be used as an approximation and a close equivalent to the Wold representation for nonlinear time series, see, e.g., \citet[][Section 4]{potter2000nonlinear} and \citet[][Section II]{jorda2005estimation}.} 

Our framework assumes that the functional form of $\bm{F}(\bullet)$ is unknown and estimated, so it is not necessarily additively separable. Even if it were, applying a nonlinear function to a Gaussian shock does not necessarily yield a random variable that follows a well-known distribution. Thus, it is not generally possible to derive closed form higher-order moments. In order to explore the predictive distributions nevertheless, one can resort to simulation-based methods.

\subsection{Predictive Simulation}\label{sec:ucondfc}
Define a vector $\bm{\Xi}$ that contains all coefficients and latent variables necessary to parameterize (\ref{eq:general-model}). At time $\tau$, the one-step ahead predictive distribution is:
\begin{equation}
    p_\tau(\bm{y}_{\tau+1} \cond \mathcal{I}) = \int p_\tau(\bm{y}_{\tau+1} \cond \mathcal{I}, \bm{\Xi}) p(\bm{\Xi} \cond \mathcal{I}) d \bm{\Xi},\label{eq:one-stepahead}
\end{equation}
where $\mathcal{I}$ denotes the information set used to infer $\bm{\Xi}$, and the subscript on $p_\tau(\bullet)$ marks the forecast origin. For typical out-of-sample exercises, $\mathcal{I} = \{\bm{y}_t\}_{t = 1}^{\tau}$, and we are interested in $p_\tau(\bm{y}_{\tau+h} \cond \{\bm{y}_t\}_{t = 1}^{\tau})$ for $h = 1, 2, \hdots,$ steps ahead \citep[see][]{geweke2010comparing}. In other cases we condition on $\mathcal{I}=\{\bm{y}_t\}_{t = 1}^{T}$, to compute scenarios in-sample for $\tau \in \{1,2,\hdots, T\}$ using $p_\tau(\bm{y}_{\tau+h} \cond \{\bm{y}_t\}_{t = 1}^{T})$ with parameters updated based on the full sample, but for initial conditions at time $\tau$. In either case (unless necessary for clarity, we omit the $\tau$ index for the forecast origin below), $p_\tau(\bm{y}_{\tau+1} \cond \mathcal{I})$ generally does not take a well-known form, and neither does the distribution of higher-order forecasts for $h\geq2$.

However, we may obtain random samples from them via simulation. This involves exploiting the fact that even though $p(\bm{y}_{\tau+1} \cond \mathcal{I})$ is unknown, $p(\bm{y}_{\tau+1} \cond \mathcal{I}, \bm{\Xi})$ takes a known form under the model in (\ref{eq:general-model}). The one-step ahead predictive distribution is:
\begin{equation}
    p(\bm{y}_{\tau+1} \cond \mathcal{I}, \bm{\Xi}^{(m)}) = \mathcal{N}(\bm{F}^{(m)}(\bm{x}_{\tau+1}), \bm{\Sigma}_{\tau+1}^{(m)}),\label{eq:preddist1}
\end{equation}
where $\bm{x}_{\tau+1} = (\bm{y}_{\tau}',\hdots,\bm{y}_{\tau-p+1}')'$. Here, $x^{(m)}$ refers to the $m$th draw of a random variable $x$ --- in most cases throughout this paper, $m$ indexes a draw from a posterior or predictive distribution when running a Markov chain Monte Carlo (MCMC) algorithm. In what follows we omit the $m$-indexing for the parameters, but stress that these computations are carried out in each sweep of the algorithm and thus account for the posterior uncertainty of parameters. For $h \geq 2$ we iterate forward, conditioning recursively on the draws for preceding horizons, by setting the predictors to $\bm{x}_{\tau+h}^{(m)}=(\bm{y}_{\tau+h-1}^{(m)\prime},\bm{y}_{\tau+h-2}^{(m)\prime},\hdots)'$, and obtain:
\begin{equation}
    p(\bm{y}_{\tau+h} \cond \mathcal{I}, \bm{y}_{\tau+1:\tau+h-1}^{(m)}, \bm{\Xi}) = \mathcal{N}(\bm{F}(\bm{x}_{\tau+h}^{(m)}), \bm{\Sigma}_{\tau+h}),\label{eq:preddisth}
\end{equation}
where $\bm{y}_{\tau+1:\tau+h-1}^{(m)}$ denotes the path of the variables from $\tau+1$ to $\tau+h-1$ and $\bm{y}_{\tau+1:\tau+h} = (\bm{y}_{\tau+1}',\hdots,\bm{y}_{\tau+h}')'$. This relates to the recursive composition discussed in the context of (\ref{eq:Fhpred}), and exploits the fact that:
\begin{equation}
    p(\bm{y}_{\tau+1:\tau+h} \cond \mathcal{I}) = \int p(\bm{y}_{\tau+1} \cond \mathcal{I},\bm{\Xi}) \prod_{j = 2}^h p(\bm{y}_{\tau+j} \cond \bm{y}_{\tau+1:\tau+j-1}, \mathcal{I},\bm{\Xi}) p(\bm{\Xi} \cond \mathcal{I}) d\bm{\Xi},\label{eq:preddistjoint}
\end{equation}
that is, the joint predictive distribution factors into the product of the conditional one-step ahead densities. Simulating the process forward, sampling from the distribution in (\ref{eq:preddisth}) across horizons $h = 1,2,\hdots,$ in each sweep of an MCMC algorithm, delivers draws from $p(\bm{y}_{\tau+1:\tau+h} \cond \mathcal{I})$ via a Monte Carlo approach.

\subsection{Conditional Forecasts}\label{sec:condfc}
As their name suggests, CFs imply an additional conditioning argument in (\ref{eq:preddistjoint}). In this context, we denote by $\mathcal{C}_{h}$ a set that defines desired restrictions at horizon $h = 1, 2, \hdots,H,$ where $H$ is the maximum forecast horizon of interest, $\mathcal{C}_{1:H} = \{\mathcal{C}_{1},\hdots,\mathcal{C}_{H}\}$ collects all restrictions over the full forecast path, and the unconditional forecast results when $\mathcal{C}_{h} = \emptyset$ for all $h$. The joint distribution of interest is $p(\bm{y}_{\tau+1:\tau+H} \cond \mathcal{I}, \mathcal{C}_{1:H})$, and can be obtained by marginalizing $p(\bm{y}_{\tau+1:\tau+H} \cond \mathcal{I}, \mathcal{C}_{1:H},\bm{\Xi})$ over the parameters via MCMC sampling, as discussed previously. 

When there are no restrictions (i.e., for unconditional forecasts), the model can be simulated forward according to the law of motion in (\ref{eq:general-model}), so that any jointly realized random path over $(\tau+1):(\tau+H)$ is consistent with the model's dynamics. In the presence of restrictions, however, the key challenge in the nonlinear context is that we cannot work directly with the joint distribution to impose them across all horizons simultaneously, as is possible in linear settings; and a decomposition like in (\ref{eq:preddistjoint}) is generally not applicable.

We propose to use a particle MCMC algorithm \citep[see][]{Andrieu2010}, adapted from the literature on nonlinear state space models, as a solution. Specifically, our approach is based on PGAS as proposed by \citet{lindsten14a}, akin to the filtering-based conditional forecast implementations of \citet{banbura2015conditional}, and capable of enforcing the restrictions discussed in more detail below on an $h$-by-$h$ basis. In the spirit of \citet{andersson2010density}, we define $\mathcal{C}_{h}$ as a stochastic restriction:\footnote{The restrictions can also be written as a regression, $\bm{r}_h = \bm{R}_h\bm{y}_{\tau+h} + \bm{\eta}_{h}$ with $\bm{\eta}_{h}\sim\mathcal{N}(\bm{0},\bm{\Omega}_h)$, which is equivalent in distribution. The regression illustrates the interpretation as a type of measurement equation. Exact (hard) restrictions can be approximated by setting the respective elements of $\bm{\Omega}_h$ to, e.g., $10^{-8}$, which has computational and numerical advantages over using exact zeroes.}
\begin{align}
    \bm{R}_h\bm{y}_{\tau+h}\sim\mathcal{N}\left(\bm{r}_h,\bm{\Omega}_h\right),\label{eq:dist-restr}
\end{align}
where $\bm{R}_h$ is an $r_h \times n$ horizon-specific matrix selecting or encoding $r_h$ restrictions on linear combinations of endogenous variables, and $\bm{r}_h$ and $\bm{\Omega}_h$ are the $r_h\times1$ mean and $r_h\times r_h$ covariance matrix of the restrictions.

\inlinehead{Structural scenarios} In many cases it is desirable to impose a mix of restrictions on future observables as well as structural shocks, for structural scenario analysis in the spirit of \citet{antolin2021structural}. This allows for imposing that specific structural shocks (rather than an unrestricted combination of them) are responsible for the respective scenario in terms of the observables. Econometrically, this can be achieved by forcing a subset of the structural shocks to their unconditional distribution (non-driving shocks), and leave the remaining ones (driving shocks) free. The latter thus may deviate from their unconditional distribution, thereby acting as the offsetting forces that yield the restrictions imposed on the observables. The same intuition can be used for computing dynamic responses to specific structural shocks, see also \citet{breitenlechner2022goes}.

As discussed above, the one-step ahead prediction of our model conditional on any preceding information up to $\tau+h-1$ can be written as $\bm{y}_{\tau+h} = \bm{\mu}_{\tau+h}^{(m)} + \bm{\epsilon}_{\tau+h}$, with $\bm{\mu}_{\tau+h}^{(m)} = \bm{F}(\bm{y}_{\tau+h-1}^{(m)}, \hdots, \bm{y}_{\tau+h-p}^{(m)}) = \bm{F}(\bm{x}_{\tau+h}^{(m)})$ and $\bm{\epsilon}_{\tau+h} \sim \mathcal{N}(\bm{0},\bm{\Sigma}_{\tau+h})$. We abstract below from time-variation in the variances to simplify notation, but note that this framework is applicable also in the presence of heteroskedastic shocks. A structural form of the model can be obtained as $\bm{H}\bm{y}_{\tau+h} = \bm{H}\bm{\mu}_{\tau+h} + \bm{u}_{\tau+h}$, with iid shocks $\bm{u}_{\tau}\sim\mathcal{N}(\bm{0}_n,\bm{I}_n)$, where $\bm{I}_n$ is an $n$-dimensional identity matrix. Then, $\bm{\epsilon}_{\tau} = \bm{H}^{-1}\bm{u}_{\tau}$, and $\bm{\Sigma} = \bm{H}^{-1}\bm{H}^{-1}{'}$.\footnote{An alternative, see also Section \ref{sec:econometrics}, is to parameterize (\ref{eq:general-model}) directly as $\bm{H}\bm{y}_t = \bm{F}(\bm{x}_t) + \bm{u}_t,~\bm{u}_t\sim\mathcal{N}(\bm{0}_n,\bm{I}_n)$, where $\bm{H}$ is nonsingular, see also \citet{arias2023macroeconomic,chan2024large}. Time-varying variances of structural shocks, or more general forms of heteroskedasticity, can be addressed by using $\bm{u}_{\tau}\sim\mathcal{N}(\bm{0}_n,\bm{S}_\tau)$, where $\bm{S}_\tau = \diag(s_{1\tau}^2,\hdots,s_{n\tau}^2)$, such that $\bm{\Sigma}_\tau = \bm{H}^{-1}\bm{S}_\tau\bm{H}^{-1}{'}$.} Both types of restrictions can be implemented with distributional assumptions about the restricted vector of endogenous variables, as $r_h = r_h^{(y)} + r_h^{(u)}$ restrictions on:
\begin{enumerate}[label=(\roman*)]
    \item {observables}: $\bm{R}^{(y)}_h\bm{y}_{\tau+h} \sim \mathcal{N}(\bm{r}^{(y)}_h, \bm{\Omega}^{(y)}_h)$, imposing $r_h^{(y)}$ restrictions;
    \item {structural shocks}: $\bm{R}^{(u)}\bm{u}_{\tau+h} \sim \mathcal{N}(\bm{r}^{(u)}_h,\bm{\Omega}^{(u)}_h)$, or equivalently, $\bm{R}^{(u)}\bm{H}\bm{y}_{\tau+h} \sim \mathcal{N}(\bm{r}^{(u)}_h + \bm{R}^{(u)}\bm{H}\bm{\mu}_{\tau+h},\bm{\Omega}^{(u)}_h)$, imposing $r_h^{(u)}$ restrictions.
\end{enumerate}
We may write these in terms of (\ref{eq:dist-restr}) by stacking both types in terms of their implications about the observables. The operator $\text{bdiag}(\bullet)$ outputs a block diagonal matrix, with the inputs arranged along the main block diagonal; we obtain $\bm{r}_h = [\bm{r}^{(y)}_h; \bm{r}^{(u)}_h + \bm{R}^{(u)}\bm{H}\bm{\mu}_{\tau+h}]$, $\bm{R}_h = [\bm{R}^{(y)}_{h}; \bm{R}^{(u)}_{h}\bm{H}]$ and $\bm{\Omega}_h = \text{bdiag}(\bm{\Omega}^{(y)}_h, \bm{\Omega}^{(u)}_h)$. Under these assumptions we obtain the joint distribution of the forecast and the restrictions at horizon $h$:
\begin{equation}
    \begin{bmatrix}
        \bm{y}_{\tau+h}\\
        \bm{r}_h
    \end{bmatrix} \sim \mathcal{N}\left(
    \begin{bmatrix}
        \bm{I}_n\\
        \bm{R}_h
    \end{bmatrix} \bm{\mu}_{\tau+h}^{(m)}, 
    \begin{bmatrix}
        \bm{\Sigma} & \bm{\Sigma}\bm{R}_h'\\
        \bm{R}_h\bm{\Sigma} & \bm{R}_h\bm{\Sigma}\bm{R}_h' + \bm{\Omega}_h
    \end{bmatrix}
    \right).\label{eq:jointfcrestr}
\end{equation}
which encodes a nonlinear state combined with a linear measurement equation. This casts conditional forecasting and structural scenario analysis as a state space problem, similar to \citet{banbura2015conditional}, and allows to apply the PGAS algorithm of \citet{lindsten14a} with a few adjustments.\footnote{Our framework is also applicable for nowcasting problems \citep[see, e.g.,][]{cimadomo2022nowcasting}.} 

\inlinehead{Particle Gibbs with Ancestor Sampling} The idea of particle Gibbs algorithms is to use hypothetical realizations, a set of particles $v = 1,\hdots,V,$ combined with a scheme to score how plausible these particles are in light of the measurements/restrictions. A key ingredient of PGAS is that we condition on a fixed trajectory for one of the particles. This reference particle is chosen as the previous draw for the respective states in the encompassing MCMC sampler, ensuring that the Markov kernel preserves the correct invariant distribution \citep[see][]{Andrieu2010}. The ``ancestor'' part of PGAS adds a resampling step for the parents of the reference trajectory, which alleviates path degeneracy by maintaining variability in the ancestral lineages, see \citet{lindsten14a} for details. This improves mixing, especially with few particles, making the algorithm computationally attractive.

The $h$-specific distribution of the forecast conditional on the restrictions is available in closed form \citep[see, e.g.,][chapter 16]{west1997bayesian} when assuming additive Gaussian errors and restrictions as in (\ref{eq:dist-restr}):
\begin{align}
    \bm{y}_{\tau+h} &\cond \bm{r}_h,\bm{R}_h,\bm{\Omega}_h,\bullet \sim \mathcal{N}(\bm{r}^{\ast}_h, \bm{\Omega}_h^{\ast})\label{eq:condfcst}\\
    \bm{r}^{\ast}_h &= \bm{\mu}_{\tau+h}^{(m)} + \bm{\Sigma}\bm{R}_h'(\bm{R}_h\bm{\Sigma}\bm{R}_h' + \bm{\Omega}_h)^{-1}(\bm{r}_h - \bm{R}_h \bm{\mu}_{\tau+h}^{(m)}),\nonumber\\
    \bm{\Omega}_h^{\ast} &= \bm{\Sigma} - \bm{\Sigma}\bm{R}_h'
    (\bm{R}_h\bm{\Sigma}\bm{R}_h' + \bm{\Omega}_h)^{-1} \bm{R}_h\bm{\Sigma}.\nonumber
\end{align}
While (\ref{eq:condfcst}) places the respective restriction only at a single horizon, this conditional Gaussian distribution can be used as an ``optimal proposal'' for particles in algorithm, in the spirit of a forward-filtering update. This allows to generate particles that satisfy the restrictions locally without being exposed to excessive weight degeneracy (the situation in which only very few, if any, particles are compatible with the restrictions). Used within an encompassing MCMC algorithm, PGAS propagates these particles in such a way that the restrictions are enforced jointly over the entire forecast path, and we obtain draws from $p(\bm{y}_{\tau+1:\tau+H} \cond \mathcal{I}, \mathcal{C}_{1:H})$.

Details about PGAS and explicit computations appear in Appendix \ref{app:technical}. We provide a brief summary of the main steps below. The initial conditions at each forecast origin $\tau$ are known, and we can generate $V-1$ candidate particles to be considered alongside the reference particle from the previous MCMC draw. We loop through:
\begin{enumerate}[label=(\roman*)]
    \item \textit{Resampling and ancestor sampling}: To only retain particles with a high probability of having generated the measurements, i.e., the restrictions in (\ref{eq:dist-restr}), we sample ancestors for the non-reference particles using weights from the previous horizon, see step (iii) below. For the reference particle, we sample an ancestor index from a categorical distribution with support $\{1,\hdots,V\}$, where the weights are proportional to the likelihood of particle $v = 1,\hdots,V,$ having generated the reference. Put simply, we obtain a random ancestry that is likely to have generated the reference path.
    \item \textit{Propagation}: We then propagate the ``surviving'' non-reference particles from the resampling step one period forward. In case there are no restrictions ($\mathcal{C}_h = \emptyset$) this is done using (\ref{eq:preddisth}). In case there are restrictions, we use (\ref{eq:condfcst}) to draw particles already conditioned on them at $h$. This avoids generating unsuitable candidate realizations and reduces the required number of particles $V$.
    \item \textit{Weighting}: We update the weights used in step (i) for the next horizon based on (\ref{eq:dist-restr}). Specifically, we compute the likelihood of the lineage of the respective particles in light of the restrictions. In case there are no restrictions, we obtain equal weights.
\end{enumerate}
Using the weights at the maximum forecast horizon $H$, we draw a particle index and generate a fully smoothed trajectory by tracing back through the stored ancestor indices. This trajectory becomes the new reference path in the next iteration. These steps are run in each sweep of the main MCMC algorithm. In some cases we require the expectation of the predictive distribution in each MCMC sweep. We obtain this moment via exploiting a backward recursion \citep[see, e.g.,][]{godsill2004monte} to obtain smoothing weights, and use the full set of particle trajectories for related computations, see also Appendix \ref{app:technical}. We provide a comparison of the precision sampler of \citet{chan2023conditional} with our PGAS approach in a linear VAR (with closed form solutions) using artificial data in Appendix \ref{app:empirics}.

\subsection{Generalized Impulse Response Functions}\label{sec:GIRFs}
Various types of IRFs are widely used for both academic and policy analysis. These types of dynamic effects can generally be defined as the difference between two forecasts, see \citet{gallant1993nonlinear,koop1996impulse}. This provides a natural link to our previous discussions on predictive distributions. Indeed, due to the recursive nonlinear structure of our model, we need to resort to a simulation-based version of the GIRF.\footnote{Simulation-based methods are often used to compute (generalized) impulse response functions in nonlinear  parametric \citep[e.g.,][]{baumeister2013time,alessandri2019financial}, or nonparametric \citep[e.g.,][]{huber2022inference,clark2025nonparametric,hauzenberger2024gaussian} models.} We consider three main variants, which are nested in the expression:
\begin{equation}
    \bm{\Delta}_{\tau} = \mathbb{E}(\bm{y}_{\tau+1:\tau+H}\cond\mathcal{C}_{1:H}^{(\tts)},\bm{x}_{\tau+1},\mathcal{I}) - \mathbb{E}(\bm{y}_{\tau+1:\tau+H}\cond\mathcal{C}_{1:H}^{(\ttb)},\bm{x}_{\tau+1},\mathcal{I}),\label{eq:GIRF}
\end{equation}
and differentiated by distinct conditioning assumptions. In line with the terminology in \citet{crump2021large}, the first expectation refers to the ``scenario'' (\texttt{s}) forecast imposed with the restrictions $\mathcal{C}_{1:H}^{(\tts)}$, and the latter is the ``baseline'' (\texttt{b}) forecast subject to $\mathcal{C}_{1:H}^{(\ttb)}$. Note that the conditioning on $\mathcal{I}$ here refers to the fact that these quantities are computed with draws for parameters conditioning on all available information; the expectation is taken at time $\tau$, as indicated by conditioning on $\bm{x}_{\tau+1}$. For later reference, define horizon specific GIRFs $\bm{\delta}_{\tau,h}$ based on the structure of $\bm{\Delta}_{\tau} = (\bm{\delta}_{\tau,1}^{\prime},\hdots,\bm{\delta}_{\tau,H}^{\prime})'$. That is, the one-step ahead forecast horizon is associated with the \textit{contemporaneous} impact of the restrictions \citep[see][for a similar timing convention]{breitenlechner2022goes}.

Equation (\ref{eq:GIRF}) nests our main variants:\footnote{Other related work includes \citet{bernanke1997systematic,hamilton2004comment,sims2006does,baumeister2012unconventional,baumeister2014real,adrian2025scenario}.} the (1) unorthogonalized GIRF (UGIRF) which can be obtained from a fully reduced form model, but which may be extended to a structural scenario analysis when additionally identifying and restricting structural shocks; the (2) structural GIRF (SGIRF) in response to a structurally identified shock imposed exclusively with restrictions on the contemporaneous ($h=1$) shocks; and, the (3) restricted GIRF (RGIRF), using an SGIRF as a baseline but systematically manipulating transmission channels in the spirit of assessing alternate policy rules, by additionally imposing restrictions on observables in a certain way.

\inlinehead{Unorthogonalized GIRF} This version is closely related to the original GIRF of \citet{koop1996impulse}. It is obtained by restricting only observables as the scenario $\mathcal{C}_{1:H}^{(\tts)}$, which can be compared to the unconditional forecast, $\mathcal{C}_{1:H}^{(\ttb)} = \emptyset$, as a baseline. Without any further restrictions, UGIRFs simply reflect a likely combination of structural shocks that drive the change in the respective observable(s), see also \citet{crump2021large}. In case one is willing to impose additional structure in the model via identifying structural shocks and $\bm{H}$, restrictions on observables can be complemented with restrictions on shocks \citep{antolin2021structural}. Specifically, one may restrict a subset of the shocks to their unconditional distribution (non-driving shocks), while the unrestricted subset (driving shocks) is allowed to deviate to deliver the structural scenario in terms of the observables. In light of the assumptions about the structural form above, this implies $\bm{r}^{(u)}_h = \bm{0}$ and $\bm{\Omega}^{(u)}_h = \bm{I}$ for the \textit{non-driving} shocks.

\inlinehead{Structural GIRF} This variant relies on orthogonalized structural economic shocks. We follow the literature \citep[see, e.g.,][]{jordalocal}, and define the SGIRF in response to the $j$th structural shock of magnitude $d$ as:
\begin{equation}
    \bm{\Delta}_{j\tau}^{(d)} = \mathbb{E}(\bm{y}_{\tau+1:\tau+H} \cond u_{j\tau+1} = d_0 + d, \bm{x}_{\tau+1}, \mathcal{I}) - \mathbb{E}(\bm{y}_{\tau+1:\tau+H} \cond u_{j\tau+1} = d_0, \bm{x}_{\tau+1}, \mathcal{I}).\label{eq:SGIRF}
\end{equation}
The parameter $d_0$ is the baseline level of the shock. Since the structural errors enter linearly in our model, choices about the baseline level do not matter and we implicitly use $d_0 = 0$ in most of our subsequent discussions. Equation (\ref{eq:SGIRF}) is a special case of (\ref{eq:GIRF}), which can be obtained by defining $\mathcal{C}_1^{(\tts)}$ with $\bm{r}_1^{(u)} = (d_0 + d)\cdot\bm{e}_j'$, and $\mathcal{C}_1^{(\ttb)}$ with $\bm{r}_1^{(u)} = d_0\cdot\bm{e}_j'$; $\bm{e}_j'$ is the $j$th column of $\bm{I}_n$. For both forecasts we use $\bm{\Omega}_1^{(u)} = \diag(1,\hdots,1,10^{-8},1,\hdots,1)$ with an approximately binding restriction in the $j$th position, $\bm{R}_1^{(u)} = \bm{I}_n$, and we have $\mathcal{C}_h = \emptyset$ for $h > 1$. The GIRF at time $\tau$ and horizon $h$ in response to the $j$th structural shock of size $d$ is referred to as $\bm{\delta}_{j\tau,h}^{(d)}$. In Appendix \ref{app:technical} we discuss an alternative computation method which can be used without running PGAS, and Appendix \ref{app:empirics} again provides an illustration and comparison in a linear context using artificial data.

\inlinehead{Restricted GIRF} The final variant we consider combines the SGIRF with restrictions on observables, so that we may impose $\mathbb{E}(\bm{R}^{(y)}_h \bm{\delta}_{\tau,h}^{(d)}) = \bm{0}_{r_h^{(y)}}$ along the desired dimensions. This approach switches off specific transmission channels of structural shocks, by partially matching the observables of the scenario forecast conditional on the shock with those of the forecasted observables in the baseline predictive distribution. \citet{breitenlechner2022goes} provide a discussion of this approach in a linear context, which allows for directly restricting the IRF. In our nonlinear setting, we need to account for varying initial conditions and work with the two conditional forecasts, as we cannot directly manipulate the IRF.

From an implementation perspective, we may obtain a draw $\bm{y}_{\tau+1:\tau+H}^{(\ttb,m)}$ from the baseline distribution $p(\bm{y}_{\tau+1:\tau+H} \cond \mathcal{C}_1^{(\ttb)}, \bm{x}_{\tau+1}, \mathcal{I})$, where $\mathcal{C}_1^{(\ttb)}$ features the impact restrictions as in (\ref{eq:SGIRF}) plus any desired restrictions on the future sequence of non-driving shocks (this is however not a necessary requirement, as otherwise all shocks may be allowed to deviate from their unconditional distribution; or, one can also force all of them to follow their unconditional distribution, see below). For computing the scenario forecast, we then augment the shock impact restriction in (\ref{eq:SGIRF}) with a restriction on the respective observable, so that $\bm{r}_h^{(y)} = \bm{R}_h^{(y)}\bm{y}_{\tau+h}^{(\ttb,m)}$, to define $\mathcal{C}_{1:H}^{(\tts)}$. Individual \textit{realized paths} from the scenario and baseline forecast distributions are identical along the restricted dimensions. The restricted GIRF then arises as the difference in \textit{expected values} as in (\ref{eq:GIRF}), which has an expected value of zero across MCMC iterations along the restricted dimensions, but this cannot strictly be enforced for each individual draw. This is because otherwise the reference trajectory in each sweep of the PGAS algorithm may not necessarily respect the restrictions.

\inlinehead{Conditional and unconditional GIRFs} One may obtain all three of these GIRFs conditional on each point in time $\tau$. In principle, one may thus consider ``time-varying'' dynamic effects of shocks for each period individually (these estimates are also referred to as ``filtered'' in the sense that they condition on the realized history up to a specific period, see \citealp{rambachan2021common}). This time variation, however, is exclusively and mechanically due to variation across initial conditions, since the functional form in the class of models we consider is potentially nonlinear but time-invariant. Unconditional (G)IRFs can be computed in various ways (e.g., by randomizing over initial conditions and averaging, in bootstrap-type approaches); see also \citet[][chapter 18]{kilian2017structural}, and \citet{gonccalves2021impulse,gonccalves2024state} for definitions. In case we require estimates of unconditional versions, our preferred approach is to compute the GIRF at each point in time, and then average out this source of randomness, $\overline{\bm{\Delta}} = \sum_{\tau = 1}^T \bm{\Delta}_{\tau}/T$. It is worth noting that under several assumptions, see \citet[][Theorem 5]{rambachan2021common}, temporally averaged GIRFs can identify a dynamic causal effect (an ``average treatment effect,'' in related terminology).

\inlinehead{Lucas critique} Conditional forecasts and the GIRF variants relate to work on evaluating counterfactual policy rules, and may be exposed to the \citeauthor{lucas1976} critique. This is the case particularly when imposing highly unusual conditioning scenarios, as opposed to mere ``modest policy interventions,'' which may leave behavioral patterns of economic agents unchanged \citep[see][]{leeper2003modest}. A recent discussion of these aspects is also provided in \citet{mckay2023can}; a related issue is that any imposed counterfactual paths on observables in this spirit require specific neutralizing \textit{future} shocks, and the restrictions hold only ex post and not necessarily in expectation. Since our framework is based on the idea of a future sequence of shocks delivering the scenario as in \citet{antolin2021structural}, similar concerns as in linear models apply. 

Two further remarks are relevant. First, by virtue of our nonparametric approach, structural shocks may locally affect relationships among variables, even if the underlying functional relationships remain constant over time. Thus, while our baseline setting is susceptible to the \citeauthor{lucas1976} critique, it is likely more robust to related concerns than linear models. Second, \citet[][see their Appendix C]{breitenlechner2024fiscal} recently recognized that the \citeauthor{lucas1976}-robust approach of \citet{mckay2023can} can be implemented in linear frameworks of the \citet{antolin2021structural} type by allowing offsetting shocks \textit{exclusively} on impact. This restriction can straightforwardly be implemented in our setup, provided the matrix $\bm{H}$ is identified. In the spirit of our discussion in Section \ref{sec:predictions}, to subsequently explore the future conditional moments of the observables stochastically, we further augmented the desired impact restriction with a restriction forcing all future structural shocks (for $h > 1$) to adhere to their unconditional distribution.

\section{Model Specification and Estimation Algorithm}\label{sec:econometrics}
\subsection{Multivariate System Estimation}\label{sec:sampling}
To estimate the model in (\ref{eq:general-model}) we rely on a conditional representation of its $n$ equations. Let $\bm{e}_i$ of size $1 \times n$ denote the $i$th row of $\bm{I}_n$, and $\bm{E}_i$ of size $(n-1) \times n$ results from deleting the $i$th row of $\bm{I}_n$. Using $\bm{y}_{-it} = \bm{E}_i \bm{y}_t$ we may write $\bm{y}_t = \bm{e}_i'y_{it} + \bm{E}_i'\bm{y}_{-it}$. Under the assumptions of (\ref{eq:general-model}), $p(y_{it}\cond \bm{y}_{-it},\bullet) \propto \exp\left\{-(\bm{e}_i\bm{\Sigma}_{t}^{-1}\bm{e}_i'y_{it}^2 - 2 y_{it}\bm{e}_i\bm{\Sigma}_t^{-1}(\bm{F}(\bm{x}_t) - \bm{E}_i'\bm{y}_{-it}))/2\right\}$, which is a Gaussian with variance $\varsigma_{it}^2 = (\bm{e}_i\bm{\Sigma}_{t}^{-1}\bm{e}_i')^{-1}$ and mean ${\mu}_{it} = \varsigma_{it}^2 (\bm{e}_i\bm{\Sigma}_t^{-1}(\bm{F}(\bm{x}_t) - \bm{E}_i'\bm{y}_{-it}))$. This distribution is equivalent to a common representation of the conditional multivariate Gaussian \citep[see, e.g.,][Section 2]{cong2017fast}. The mean can alternatively be written as ${\mu}_{it} = f_i(\bm{x}_t) - \varsigma_{it}^2(\bm{e}_i\bm{\Sigma}_{t}^{-1}\bm{E}_i')(\bm{y}_{-it} - \bm{E}_i \bm{F}(\bm{x}_t))$, and we define $\tilde{\mu}_{it} = -\varsigma_{it}^2(\bm{e}_i\bm{\Sigma}_{t}^{-1}\bm{E}_i')(\bm{y}_{-it} - \bm{E}_i \bm{F}(\bm{x}_t))$, i.e., $\mu_{it} = f_i(\bm{x}_t) + \tilde{\mu}_{it}$.\footnote{Related papers often either use a mapping between the structural and reduced form of the VAR to enable equation-by-equation estimation \citep[see, e.g.,][]{hauzenberger2024gaussian}, or rely on factor models for the reduced form errors \citep[see, e.g.,][]{clark2021tail}. These approaches come with computational and inferential advantages and disadvantages. The former is simple to implement but requires parameterizing a structural form, which may cause issues such as inadvertently (instead of purposefully to achieve structural identification) breaking order-invariance of the equations. The latter allows for order-invariant inference but gives rise to the usual identification challenges of factor models. Our approach uses an order-invariant reduced form model for estimating the conditional mean relationships \citep[order-variant specifications may be consequential for conditional forecasts, see also][for discussions]{arias2023macroeconomic,chan2024large}.} The $i$th equation of the multivariate model in regression form, conditional on all other equations, is then given by:
\begin{equation}
    (y_{it} - \tilde{\mu}_{it}) = f_i(\bm{x}_t) + u_{it}, \quad u_{it} \sim \mathcal{N}(0,\varsigma_{it}^2),\label{eq:singleeq}
\end{equation}
which can be used in a Gibbs sampler to update the conditional mean relationships by looping through equations $i = 1,\hdots,n$. This approach is similar to the one of \citet{esser2024seemingly} and allows to treat each equation of the multivariate system individually, conditional on all other equations, which is computationally quick.

\subsection{Bayesian Additive Regression Trees}\label{sec:BART}
The approach we discuss in Section \ref{sec:predictions} works with any implementation of multivariate models with additive and jointly Gaussian errors. That is, assuming a linear functional form for $\bm{F}(\bm{x}_t)$ combined with suitable priors results in a standard Bayesian VAR (BVAR).\footnote{When we consider linear versions of our model for comparisons, we implement this setting with $\bm{F}(\bm{x}_t) = \bm{A}\bm{x}_t$ where $\bm{A}$ is an $n\times k$ matrix of reduced form VAR coefficients. We assume a horseshoe prior with a single global shrinkage component on these parameters, see also \citet{hauzenberger2024bookchapter}; (\ref{eq:singleeq}) can be used to update the VAR coefficients equation-by-equation from textbook Gaussian posteriors.} In case we treat $\bm{F}(\bm{x}_t)$ nonparametrically, several options are available. Due to its versatility and established favorable empirical properties we mentioned earlier, we use BART to approximate the equation-specific functions in our applied work. That is, we consider a sum of $s = 1,\hdots,S,$ tree functions $\ell_{is}(\bm{x}_t\cond\mathcal{T}_{is}, \bm{\mathrm{m}}_{is})$ such that $f_i(\bm{x}_t) \approx \sum_{s=1}^{S}\ell_{is}(\bm{x}_t\cond\mathcal{T}_{is}, \bm{\mathrm{m}}_{is})$ where $\mathcal{T}_{is}$ are regression trees and $\bm{\mathrm{m}}_{is}$ is a vector of terminal node parameters (which serve as fitted values). Instead of having a single but complex tree, BART is akin to ensemble methods, and uses a sum of many simple trees (``weak learners''), which has been shown to work well.

Using BART requires an algorithm that estimates splitting variables and thresholds for which we specify suitable priors that together yield $p(\mathcal{T}_{is})$; we further need a prior on the terminal node parameters $p(\bm{\mathrm{m}}_{is}\cond \mathcal{T}_{is})$. Our setup follows \citet{chipman2010bart} and we first define the probability that a tree ends at a specific node at depth $d = 0,1,2,\hdots,$ as $\alpha/(1+d)^\beta$, with $\alpha\in(0,1)$ and $\beta\in\mathbb{R}^{+}$. This prevents trees from getting overly complex and provides regularization (here, we rely on the default values $\alpha=0.95$, and $\beta=2$, which perform well across many datasets). For the splitting variables, we choose a uniform prior. This implies that each predictor is equally likely to be selected as a splitting variable. We further assign a uniform prior to all thresholds within the splitting rules, based on the range of the respective splitting variable.

Next we specify the prior for the terminal node parameters. On these parameters $\bm{\mathrm{m}}_{is,l}$, for $l = 1,\hdots,\#\text{TN}_{is}$, where $\#\text{TN}_{is}$ denotes the number of terminal node parameters of tree $s$ in equation $i$, we impose independent conjugate Gaussian priors that are symmetric across trees and identical for all terminal nodes. As suggested by \citet{chipman2010bart} the moments of these priors are chosen in a data-driven manner, such that $95$\% of the prior probability lies in the interval $(\min(\bm{y}_i),\max(\bm{y}_i))$, where $\bm{y}_i = (y_{i1},\hdots,y_{iT})'$, and such that shrinkage increases the more trees $S$ are chosen for estimation. We choose $S=250$ trees which has been shown to work well for typical macroeconomic time series applications \citep[see, e.g.,][]{huber2023nowcasting}. Additional details are provided in Appendix \ref{app:technical}.

\subsection{Other Priors and Sampling Algorithm}\label{sec:otheraspects}
Inspired by \citet{carriero2021addressing}, we assume that $\bm{\Sigma}_t = s_t^2 \bm{\Sigma}$ --- the covariance structure varies proportionally over time and $s_t$ is used to capture outliers. Our prior setup for the constant part of the covariance matrix follows \citet{esser2024seemingly}. Specifically, we use a hierarchical inverse Wishart prior $\bm{\Sigma} \cond \{a_i\}_{i=1}^n \sim \mathcal{W}^{-1}(s_0, \bm{S}_0),$ where $s_0 = \nu + n - 1$, $\bm{S}_0 = 2 \nu \cdot \diag(1/a_1,\hdots,1/a_n)$ and $a_i\sim\mathcal{G}^{-1}(1/2,1/A_j^2)$ for $i = 1,\hdots,n,$ and a fixed scale parameter $A_j > 0$. Setting $\nu = 2$ implies a comparatively uninformative prior about the implied correlation structure, different from fixed-hyperparameter versions of this prior which has a tendency to overshrink. If applicable, we assume that $\Pr(s_t = 1) = 1 - \mathfrak{p}$ and $\Pr(s_t \sim \mathcal{U}(2,\overline{\mathfrak{s}})) = \mathfrak{p}$, where $\mathcal{U}(2,\overline{\mathfrak{s}})$ is a discrete uniform distribution with (integer) support between $2$ and $\overline{\mathfrak{s}} = 6$ and $\mathfrak{p}\sim\mathcal{B}(a_\mathfrak{p},b_\mathfrak{p})$ is the probability of observing an outlier. We set $a_\mathfrak{p} = 1$, $b_\mathfrak{p} = 50$ which a priori implies about $2$\% of the observations are outliers. Alternative and more flexible stochastic volatility specifications are available in this context \citep[e.g.,][]{chan2023comparing}.\footnote{Another computationally attractive variant is to assume $\bm{\Sigma}_t = \bm{S}_t\bm{\Sigma}\bm{S}_t$ where $\bm{S}_t = \diag(s_{1t},\hdots,s_{nt})$, which allows for variable-specific outlier detection using the same setup as in Section \ref{sec:otheraspects}. These assumptions allow to invert $\bm{\Sigma}$ only once instead of having to invert $\bm{\Sigma}_t$ for all $t$ to compute the moments in (\ref{eq:singleeq}).} We obtain draws from the joint posterior of our model using a fairly straightforward MCMC algorithm. Details are provided in Appendix \ref{app:technical}.

\section{Empirical Applications}\label{sec:application}
We employ the proposed framework in three related yet distinct applications. First, we use the annual stress test scenarios conducted by the Federal Reserve System and compute forecasts conditional on multiple observables. Second, we study the implications of varying financial conditions on tail risks of output growth, inflation, and employment. Third, we identify a US-based financial shock and gauge the role of spillovers and spillbacks.\footnote{We sometimes compare a \textit{hom}oskedastic (\texttt{BART-hom}, $s_t = 1$ for all $t$) and \textit{het}eroskedastic BART (\texttt{BART-het}) with a \textit{het}eroskedastic BVAR (\texttt{BVAR-het}), featuring the outlier specification. The linear BVAR serves for comparisons as it is a popular workhorse model in related contexts \citep[e.g.,][]{crump2021large}; due to our sample featuring the Covid-19 pandemic, we disregard its homoskedastic version.}

\subsection{Stress Testing Scenarios for the US Economy}\label{sec:dfst}
In our first application, we conduct a scenario analysis for the US economy inspired by the 2025 version of the \href{https://www.federalreserve.gov/publications/dodd-frank-act-stress-test-publications.htm}{\textit{Dodd-Frank Act}} (DFA) stress test assumptions. This annual exercise is conducted and published by the Board of Governors of the Federal Reserve System. Details about the underlying dataset are provided in Appendix \ref{app:empirics}. The information set features about $25$ broad variables (capturing economic activity, labor market, prices, housing and the financial sector). We estimate our models using quarterly data ranging from 1976Q1 to 2024Q4, and subsequently consider a baseline and adverse scenario for the period from 2025Q1 to 2027Q4. These scenarios are imposed via constraints on the path of the unemployment rate (\texttt{UNRATE}), CPI inflation (\texttt{CPIAUCSL}), and $10$-year government bond yields (\texttt{GS10}), inspired by \citet{chan2023conditional}. We set the tightness of the restrictions in $\bm{\Omega}_{1:H}$ using the marginal variances on the diagonal of $\bm{\Sigma}$ in each MCMC draw.

\begin{figure}[t]
    \centering
    \includegraphics[width=\linewidth]{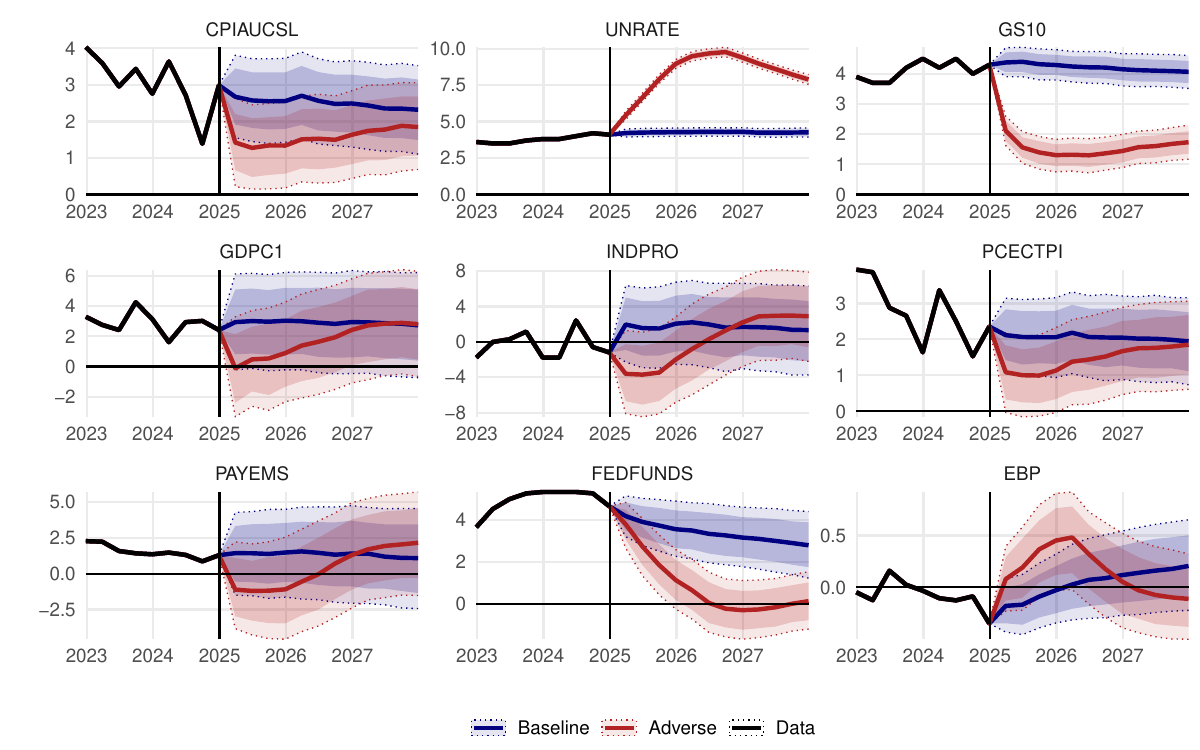}
    \caption{Conditional forecasts with \texttt{BART-het} for selected variables.}\vspace*{-0.25cm}
    \caption*{\footnotesize \textit{Notes}: Posterior median alongside 50/68 percent credible sets. Restricted variables: consumer price inflation (\texttt{CPIAUCSL}), unemployment rate (\texttt{UNRATE}), $10$-year government bond yields (\texttt{GS10}); unrestricted variables: Real GDP (\texttt{GDPC1}), industrial production (\texttt{INDPRO}), personal consumption expenditure inflation (\texttt{PCECTPI}), payroll employment (\texttt{PAYEMS}), federal funds rate (\texttt{FEDFUNDS}) and excess bond premium (\texttt{EBP}).}
    \label{fig:cf_dfst_bart_het}
\end{figure}

\begin{figure}[ht]
    \begin{subfigure}{\linewidth}
    \centering
        \caption{\texttt{BART-het}}
        \includegraphics[width=\linewidth]{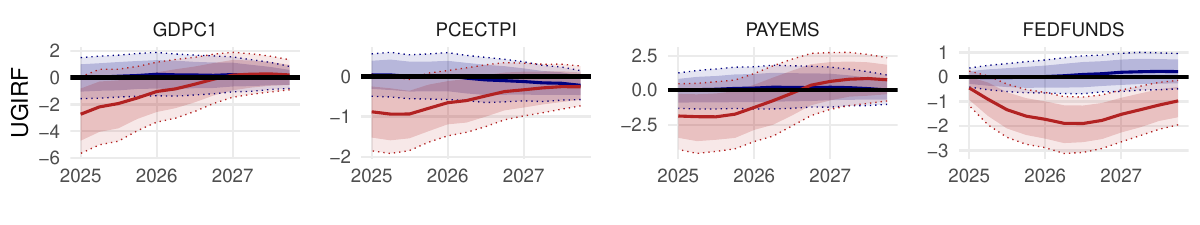}
    \end{subfigure}
    \begin{subfigure}{\textwidth}
    \centering
        \caption{\texttt{BART-hom}}
        \includegraphics[width=\linewidth]{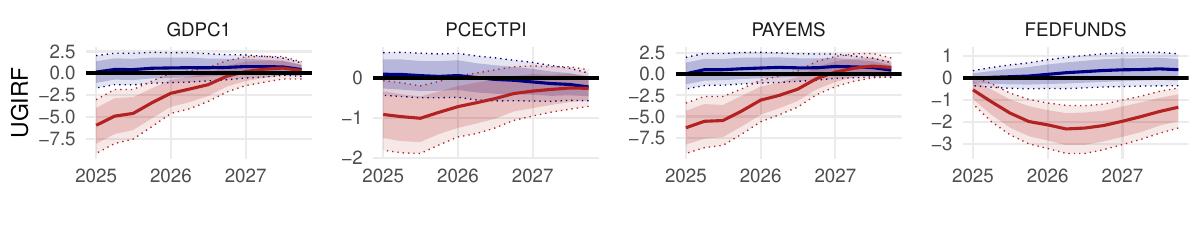}
    \end{subfigure}
    \begin{subfigure}{\textwidth}
    \centering
        \caption{\texttt{BVAR-het}}
        \includegraphics[width=\linewidth]{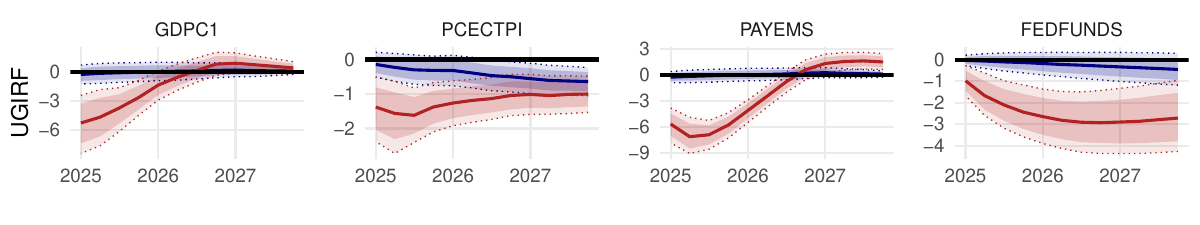}
    \end{subfigure}
    \caption{Unorthogonalized GIRF (UGIRF, scenario minus unconditional forecast) across model specifications for selected variables.}\vspace*{-0.25cm}
    \caption*{\footnotesize \textit{Notes}: Posterior median alongside 50/68 percent credible sets; adverse and baseline scenarios in red and blue, respectively. Restricted variables: consumer price inflation (\texttt{CPIAUCSL}), unemployment rate (\texttt{UNRATE}), $10$-year government bond yields (\texttt{GS10}); unrestricted variables: Real GDP (\texttt{GDPC1}), industrial production (\texttt{INDPRO}), personal consumption expenditure inflation (\texttt{PCECTPI}), payroll employment (\texttt{PAYEMS}), federal funds rate (\texttt{FEDFUNDS}) and excess bond premium (\texttt{EBP}).}
    \label{fig:cf_dfst_bart_ugirf}
\end{figure}

The posterior median forecasts and $50$ and $68$ percent posterior credible sets, obtained with \texttt{BART-het} (conditional forecast distributions for other specifications are in Appendix \ref{app:empirics}), for the restricted and selected unrestricted variables, are shown in Figure \ref{fig:cf_dfst_bart_het}: real GDP (\texttt{GDPC1}), industrial production (\texttt{INDPRO}), personal consumption expenditure (PCE) inflation (\texttt{PCECTPI}), payroll employment (\texttt{PAYEMS}), federal funds rate (\texttt{FEDFUNDS}) and the \citet{gilchrist2012credit} excess bond premium (\texttt{EBP}). The baseline scenario draws from the consensus projections from 2025 \textit{Blue Chip Financial Forecasts} and \textit{Blue Chip Economic Indicators}; the adverse scenario is characterized by a recession. Figure \ref{fig:cf_dfst_bart_ugirf} displays corresponding UGIRFs across model specifications, which are computed as the indicated scenario minus the unconditional forecast. 
  
The unconditional forecasts from \texttt{BART-het} approximately coincide with the baseline scenario (the UGIRF credible sets cover $0$ in most cases). For the adverse scenario, a different picture emerges --- the scenario forecasts differ significantly from the unconditional forecasts. There is an immediate downturn of economic activity and a reduction in payroll employment. Financial conditions tighten initially but tend to improve subsequently, partially through a monetary easing response by the central bank as reflected in the policy rate. In addition, the assumed trajectory of the conditioning variables results in a modestly disinflationary episode that vanishes by $2027$. 

The conditional forecast distributions across model specifications are similar for most variables apart from output growth, industrial production, and employment. For \texttt{BART-het}, the magnitude of the peak contraction reduces by about half. This can be explained by noting that the algorithm decides to classify several observations (that are otherwise informative about directional movements of variables when assuming homoskedasticity) as outliers. We note that BART, due to the way how tree-based approaches fit data, is capable of dealing with outliers and heteroskedastic data features in the conditional mean function by design \citep[see][for discussions]{huber2023nowcasting,clark2021tail}, even when assuming constant variances. But in this case several observations are classified as noise rather than signal. For linear models, the forecast trajectories are smoother, and an overshooting behavior is noticeable around $2027$. Moreover, the response of the federal funds rate is somewhat more persistent, which may be due to nonlinearities arising at the effective lower bound which the BVAR fails to capture.

\FloatBarrier
\subsection{Financial Conditions in the US and Tail Risk Scenarios}\label{sec:nfcitails}
In the next empirical application, we restrict our sample to 1976Q1--2017Q4 and consider the period from 2018Q1 until 2019Q1 as a laboratory to assess nonlinearities between economic variables and financial conditions. For this application we use \texttt{BART-hom} (since the pandemic observations are excluded). We investigate nonlinear patterns of macroeconomic risk, which, following the ``growth-at-risk'' approach of \citet{adrian2019vulnerable}, is defined as the predictive quantiles of some variable of interest at a pre-defined probability level (in line with value-at-risk, VaR, in finance). We pick this period because the information set already contains the global financial crisis (and the model thus had the opportunity to learn from this severe financial episode), and because this ``holdout sample'' otherwise coincides with a comparatively eventless period. We impose hard constraints on the $h = 1$ value of the National Financial Conditions Index (NFCI) and trace the effects of these scenarios on several macroeconomic variables.

\begin{figure}[t!]
    \includegraphics[width=\linewidth]{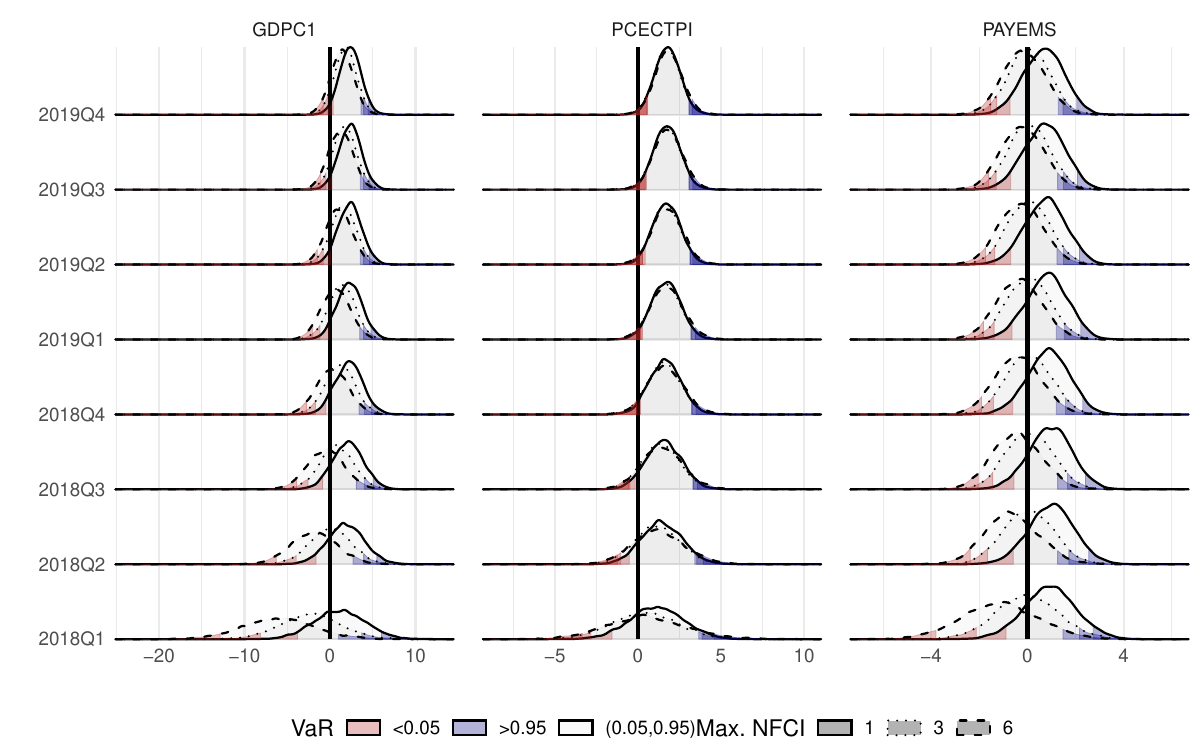}
    \includegraphics[width=\linewidth]{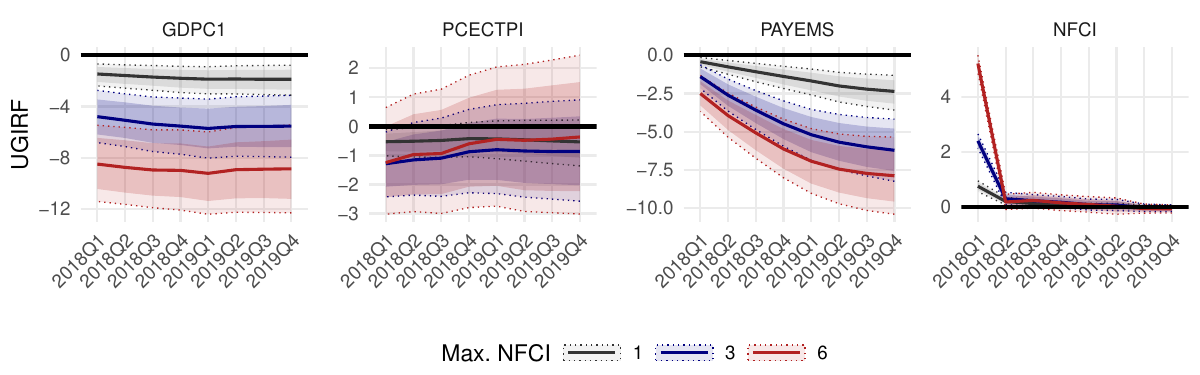}
    \caption{Conditional forecast distributions for selected variables and macroeconomic value-at-risk (VaR) and unorthogonalized GIRFs for different scenarios of financial stress.}\vspace*{-0.25cm}
    \caption*{\footnotesize \textit{Notes}: ``Max. NFCI'' refers to the maximum value of the NFCI for the scenarios, with moderate (1), severe (3, comparable to the global financial crisis), and extreme stress (6). Sampling period 1976Q1 to 2017Q4, hard restriction applies 2018Q1. Posterior median alongside 50/68 percent credible set. Variables: Real GDP (\texttt{GDPC1}), personal consumption expenditure inflation (\texttt{PCECTPI}), payroll employment (\texttt{PAYEMS}), national financial conditions index (\texttt{NFCI}). Distribution of average growth rates from 2018Q1 until the quarter indicated on the y-axis in the upper panel. Cumulated UGIRFs for all variables except the NFCI.}
    \label{fig:cf_nfci_bart}
\end{figure}

The scenarios are defined to reflect an increase of the NFCI by approximately $1$, $3$ and $6$ unconditional standard deviations (SDs, reflecting tighter financial conditions) in 2018Q1, i.e., in $\mathcal{C}_1$, which we implement by placing these values as hard restriction on the NFCI in that quarter. From 2018Q2 onward we leave the future unrestricted, i.e., $\mathcal{C}_h = \emptyset$ for $h > 1$. We investigate growth-at-risk (quantiles of real GDP), inflation-at-risk (quantiles of PCE inflation) and labor-at-risk (quantiles of growth in payroll employment) as our objects of interest \citep[see][for related papers]{adams2021forecasting,pfarrhofer2022modeling,clark2024investigating,lopez2024inflation}. The CF distributions (density estimates), for average growth rates $\overline{\bm{y}}_{T+h} = \sum_{j=1}^h\bm{y}_{T+j} / h$, are shown in Figure \ref{fig:cf_nfci_bart}. In the lower panel, we show the difference between the conditional scenario distributions and the unconditional one, which yields the UGIRF (cumulated for all variables except the NFCI).

The different NFCI scenarios for 2018Q1 shift the predictive distributions. In all cases, the economy contracts which is reflected in a decrease of real GDP growth and payroll employment, and the simulated shock has a modestly disinflationary effect. While the upper tails of the distributions (upside risk) remain comparatively stable, downside risk as measured by the lower quantiles increases significantly for all considered variables (the red shaded VaR $<0.05$ moves strongly leftwards), and there are some visible asymmetries. While the moderate NFCI scenario (max. NFCI $=1$) results in growth-at-risk for the $5$th percentile at about $-5$ percent, the severe (max. NFCI $=3$) and extreme (max. NFCI $=6$) stress scenarios yield $-7$ and $-12.5$ percent, respectively. This finding is also present for inflation-at-risk and labor-at-risk. 

The resulting predictive distributions exhibit non-Gaussian features, chief among them being heavy tails and skewness \citep[see also][for a related but simplified scenario analysis in this context]{clark2021tail}. In addition, there are hints of multimodality as the assumed values for the NFCI in 2018Q1 turn more extreme, which relates to the discussions in \citet{adrian2021multimodality}. These features can arise --- even in one-step ahead predictions and for a single restricted period --- due to, for example, the initial conditions of the nonlinear unconditional mean function at the forecast origin.
\FloatBarrier

\subsection{Spillovers and Spillbacks of US Financial Shocks}\label{sec:finshock}
In our final application, we estimate the effect of a financial shock in the US and trace its effects through the domestic economy, but also capture spillovers and spillbacks to and from other economies. We use an adjusted dataset in this case, which drops several of the domestic indicators, but adds bilateral exchange rates alongside real GDP for the EA and the UK. To identify the financial shock, we place timing-restrictions on the contemporaneous impulse responses. This is operationalized with a specific ordering of the quantities in the vector $\bm{y}_t$ --- we structure this vector such that all slow moving domestic and foreign macroeconomic variables come first (which imposes zero restrictions on impact). These variables are then followed by the EBP, and all fast moving variables such as those capturing the financial economy. We then use a Cholesky decomposition of the form $\bm{\Sigma} = \bm{P}\bm{P}'$ where $\bm{P}$ is lower triangular. That is, $\bm{H}^{-1} = \bm{P}$, and we orthogonalize the structural shocks different to our previous applications. The orthogonalized innovation of the EBP equation is interpreted as the financial shock, similar to \citet{gilchrist2012credit,barnichon2022effects}. In a multicountry context, \citet{huber2024asymmetries} use an identification scheme identical to ours.

\begin{figure}[!ht]
    \centering
    \includegraphics[width=\linewidth]{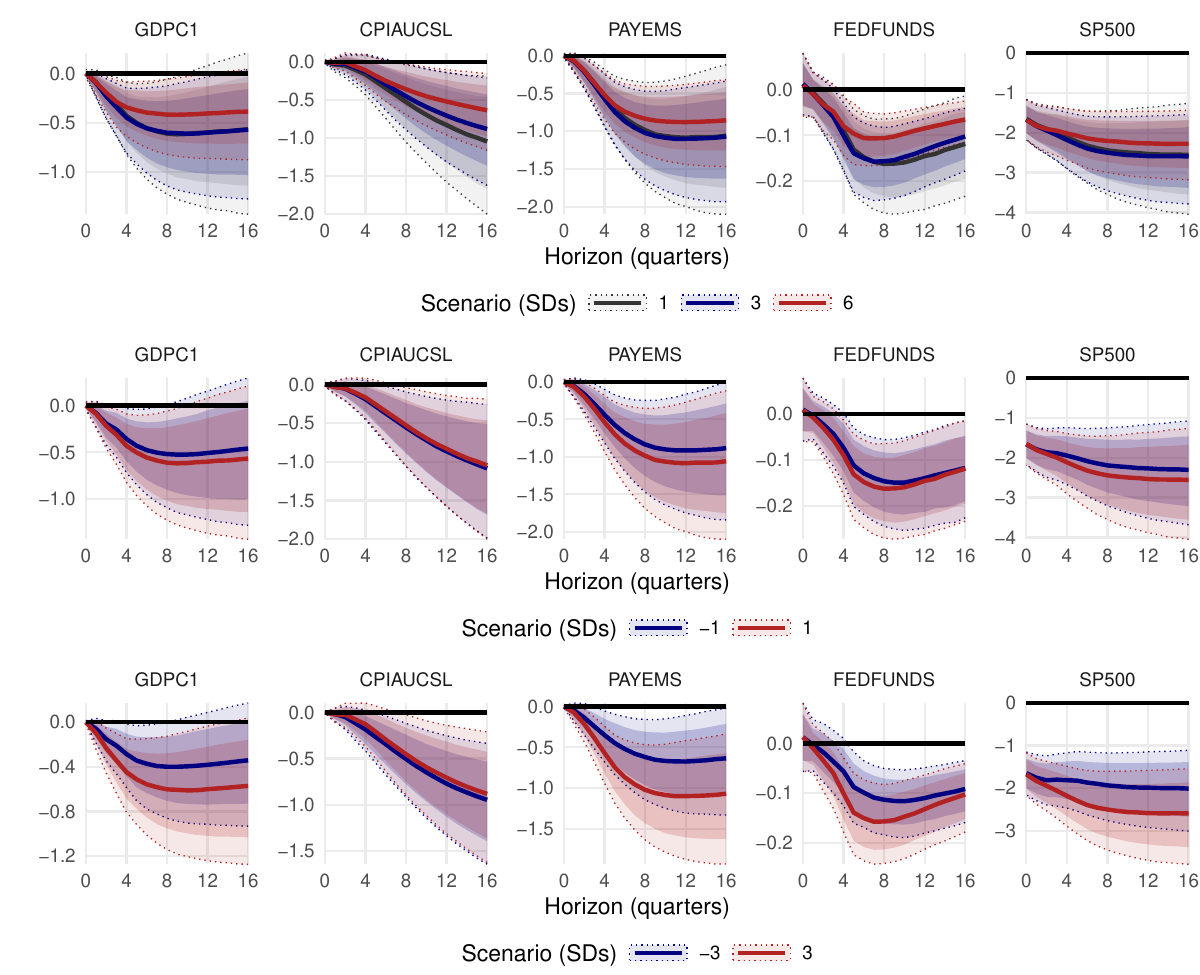}
    \caption{Structural GIRFs for selected variables to a financial shock in the US, comparing asymmetries due to size and sign of the shocks.}\vspace*{-0.25cm}
    \caption*{\footnotesize \textit{Notes}: Posterior median alongside 50/68 percent credible sets. Cumulated responses for variables in differences and levels for all other variables. ``Horizon'' refers to periods after impact of the shock. Variables: Real GDP (\texttt{GDPC1}), consumer price inflation (\texttt{CPIAUCSL}), payroll employment (\texttt{PAYEMS}), federal funds rate (\texttt{FEDFUNDS}) and S\&P500 index (\texttt{SP500}).}
    \label{fig:girfs_asym}
\end{figure}

We use $d_0 = 0$ and simulate different shock sizes and signs with $d \in \{-3,-1,1,3,6\}$. In contrast with conventional linear frameworks, our approach allows to assess nonlinearities of higher-order responses with respect to different signs and magnitudes of a proportional shock impact with GIRFs. Such asymmetries and related nonlinearities have recently gained attention both in a VAR and local projection context, see, e.g., \citet{mumtaz2022impulse,carriero2023shadow,forni2024nonlinear,hauzenberger2024machine}. While our framework allows to compute dynamic responses for each period in our sample, we focus on time averages in the results that follow. Further, we rescale all SGIRFs by computing $\bm{\delta}_{\tau,h}^{(d)} / d$ so that all responses shown below reflect a $1$ SD financial shock \citep[see also][]{gonccalves2024state,kolesar2024dynamic}. Note that for linear VARs, such scaling yields identical IRFs for all shock sizes. For nonlinear models, this is not necessarily the case and allows for a visual inspection of asymmetries. Selected variables are shown in Figure \ref{fig:girfs_asym}. The rows in the figure show different subsets of the same results, structured such that the shocks of different signs and sizes can be compared with ease.

Starting with the first row, we find that the size of the financial shock causes limited asymmetries in responses for the indicated variables. Financial shocks of different sizes rather symmetrically decrease real GDP and payroll employment. Interestingly, the effects of very large shocks increase slightly less than proportionally. Peak effects occur about two years after impact of the shock. In addition, the shock puts a persistent downward pressure on prices and leads to a decline in the federal funds rate which peaks at about $-15$ basis points after around a year. Notably the federal funds rate does not react on impact, different to stock returns which immediately decline by about $1.5$ percent during the quarter when the shock materializes. Qualitatively and in terms of magnitudes, these estimates are roughly in line with the previous literature. The US-based financial shock spills over to the other economies and leads to contractionary effects in terms of real GDP, see Appendix \ref{app:empirics}.

Having established that the size of the financial shock does not seem to matter much, the second and third row zoom into sign asymmetries. The $1$ SD US-based financial shock does not result in significant sign asymmetries (the posterior distributions overlap for the most part). Turning to the final row, this clearly differs for larger sized shocks of different signs. For these GIRFs that show the responses to a positive (adverse) and negative (benign) $3$ SD shock, asymmetries are visible for most variables. In particular, we find that the negative effect on payroll employment is almost twice as large for adverse shocks, and the Federal Reserve responds more strongly to adverse financial shocks, as measured by the much stronger shift in the federal funds rate. These findings corroborate previous evidence \citep[see, e.g.,][]{forni2024nonlinear,hauzenberger2024machine}.

\begin{figure}[t]
    \centering
    \includegraphics[width=\linewidth]{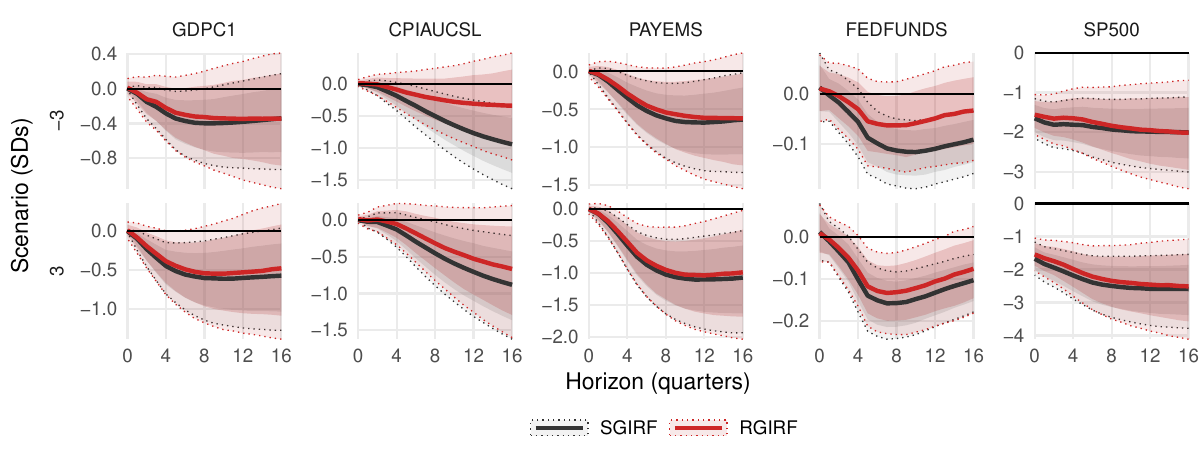}
    \caption{Structural and restricted GIRFs for selected variables.}\vspace*{-0.25cm}
    \caption*{\footnotesize \textit{Notes}: The restricted case assumes that the US financial shock does not spill over to non-domestic variables. Posterior medians alongside 50/68 percent credible sets. Cumulated responses for variables in differences and levels for all other variables. ``Horizon'' refers to periods after impact of the shock. Variables: Real GDP (\texttt{GDPC1}), consumer price inflation (\texttt{CPIAUCSL}), payroll employment (\texttt{PAYEMS}), federal funds rate (\texttt{FEDFUNDS}) and S\&P500 index (\texttt{SP500}).}
    \label{fig:girfs_diff}
\end{figure}

We next explore the role of international variables in the domestic transmission of the US shock. For this purpose, besides SGIRFs, we consider an alternative analysis where non-domestic transmission channels are switched off --- we investigate how international channels affect the domestic transmission of shocks originating in the US. That is, we impose the restriction that foreign variables (real GDP in the EA and UK, and exchange rates) do not respond to the financial shock in the US in this counterfactual, thereby simulating a scenario where the financial shock is confined to the domestic economy without any real or financial spillovers (or spillbacks); see \citet{breitenlechner2022goes} for a monetary application in this context. We restrict all domestic shocks to their unconditional distribution, and use the non-domestic ones as driving shocks to compute the RGIRFs.

The results are displayed in Figure \ref{fig:girfs_diff}. The upper panels show the SGIRFs (those shown and discussed in the context of Figure \ref{fig:girfs_asym}) and ``no spillovers'' RGIRFs. Two key findings are worth reporting. First, for the most part, ruling out spillovers and spillbacks does not significantly alter the dynamic responses after an adverse shock. It is worth mentioning, however, that restricting the international transmission leads to slightly smaller effects on average. Second, international transmission channels appear to matter for inflation dynamics, and especially so for benign financial shocks. The response of inflation turns insignificant in this case, which is also associated with a less forceful action by the central bank as captured in more muted response of the policy rate.

\FloatBarrier
\section{Conclusions}\label{sec:conclusions}
This paper presents a unified methodology for conducting scenario analysis in multivariate macroeconomic settings, accommodating nonlinearities and unknown functional forms of conditional mean relationships. These methods are applicable to traditional nonlinear frameworks, such as variants of threshold or time-varying parameter models, but also to more recently developed models incorporating Bayesian machine learning. Our framework addresses some limitations of linear and parametric models in generating various types of counterfactual analyses and is suitable for large macroeconomic datasets.

The empirical applications, using Bayesian additive regression trees as an example of nonparametric modeling of the conditional mean function, underscore the role of nonlinearities in shaping macroeconomic dynamics. For instance, a scenario analysis based on Federal Reserve stress test assumptions reveals differences between linear and nonlinear models in forecasting economic contractions and recoveries. Similarly, in a growth-at-risk application we measure nonlinear macroeconomic risks under financial stress. Finally, an analysis of financial spillovers reveals asymmetries in the transmission of shocks and the influence of international linkages on domestic outcomes.

{\setstretch{1.2}\putbib}
\end{bibunit}

\normalsize\clearpage\doublespacing
\begin{appendices}
\begin{center}
{\LARGE\sffamily\textbf{Online Appendix:\\\titletext}}
\end{center}

\setcounter{page}{1}
\setcounter{section}{0}
\setcounter{equation}{0}
\setcounter{figure}{0}
\setcounter{table}{0}
\setcounter{footnote}{0}

\renewcommand\thesection{\Alph{section}}
\renewcommand\theequation{\Alph{section}.\arabic{equation}}
\renewcommand\thetable{\Alph{section}.\arabic{table}}
\renewcommand\thefigure{\Alph{section}.\arabic{figure}}
\renewcommand\theequation{\Alph{section}.\arabic{equation}}

\begin{bibunit}
\section{Technical Details}\label{app:technical}
\subsection{Particle Gibbs with Ancestor Sampling}
We run PGAS in each iteration $m$ of our MCMC sampler, so uncertainty around model parameters is accounted for (see also our discussion in the context of unconditional forecasts in Section \ref{sec:ucondfc}); we omit the corresponding index unless it is needed for clarity. A complete sweep requires the following steps. For horizon $h = 1$, we have:
\begin{enumerate}[leftmargin =*, label=(\roman*)]
    \item The initial conditions are known and given by $\bm{x}_{\tau+1} = (\bm{y}_\tau',\hdots,\bm{y}_{\tau-p+1}')'$, that is, $\bm{\mu}_{\tau+1} = \bm{F}(\bm{x}_{\tau+1})$. We generate $V-1$ candidate particles, $\bm{y}_{\tau+1}^{(v)}$, from the unconditional predictive distribution in case $\mathcal{C}_1 = \emptyset$, and using (\ref{eq:condfcst}) in case restrictions are present. The $V$th particle serves as the fixed reference trajectory, and is set to the previous MCMC draw at iteration $m-1$, i.e., $\bm{y}_{\tau+1}^{(V)} = \bm{y}_{\tau+1}^{(m-1)}$. For later reference in the context of $h > 1$, we define $\bm{x}_{\tau+2}^{(v)} = (\bm{y}_{\tau+1}^{(v)}{'},\bm{y}_{\tau}',\hdots)'$; we discuss the required ancestry tracking for generic horizons below.
    \item The next step is to compute weights based on the measurement equation. When no restrictions are imposed we trivially obtain equal weights --- all particles are equally compatible because there are no ``measurements.'' In the presence of restrictions the unnormalized weights are generally given by $\tilde{w}_{1}^{(v)} \propto \mathcal{N}(\bm{r}_1; \bm{R}_1 \bm{\mu}_{\tau+1}, \bm{R}_1\bm{\Sigma}\bm{R}_1' + \bm{\Omega}_1)$, but since the initial conditions are fixed and $\bm{\mu}_{\tau+1}$ is the same for all particles, we obtain equal weights in either case.
\end{enumerate}
At each horizon $h = 1,2,\hdots$, we generally compute the normalized weights as
\begin{equation*}
w_{h}^{(v)} \;=\; 
\frac{\exp\!\left(\log \tilde{w}_{h}^{(v)} - \max_{j} \log \tilde{w}_{h}^{(j)}\right)}
{\sum_{j=1}^V \exp\!\left(\log \tilde{w}_{h}^{(j)} - \max_{j'} \log \tilde{w}_{h}^{(j')}\right)}.
\end{equation*}
For all subsequent horizons $h > 1$, the following steps are required:
\begin{enumerate}[leftmargin =*, label=(\roman*)]
    \item We first resample ancestor indices, $a_h^{(v)} \in \{1,\hdots,V\}$, for non-reference particles (for $v = 1,\hdots,V-1$) from a categorical distribution, $a_h^{(v)} \sim \text{Cat}(w_{h-1}^{(1)},\hdots,w_{h-1}^{(V)})$. This step serves to retain particles with a high probability of having generated the measurements (in our case, restrictions) from the previous horizon.
    
    For the reference particle $v = V$, the respective index is obtained with ancestor sampling. Specifically, we compute $\pi_h^{(v)} = \log\tilde{w}_{h-1}^{(v)} + \log \mathcal{N}(\bm{y}_{\tau+h}^{(V)}; \bm{F}(\bm{x}_{\tau+h}^{(v)}), \bm{\Sigma})$. We apply a weight-normalization analogous to the one above, to obtain $\overline{w}_{h}^{(v)}$ using $\pi_h^{(v)}$. The ancestor index for the reference trajectory is sampled from $a_h^{(V)} \sim \text{Cat}(\overline{w}_{h}^{(1)},\hdots,\overline{w}_{h}^{(V)})$. This step reweights the candidate parents of the reference trajectory with their probability of generating the reference.
    \item In the next step, we propagate the non-reference particles based on their parent index to keep the lineage consistent also in view of the lag structures. We obtain $\bm{\mu}_{\tau+h}^{(v)} = \bm{F}(\bm{x}_{\tau+h}^{(a_h^{(v)})})$. If $\mathcal{C}_h = \emptyset$, we produce samples from the unconditional one-step ahead distributions; in case there are restrictions, we use (\ref{eq:condfcst}) to draw particles that are consistent with these restrictions. For the reference particle, we again set $\bm{y}_{\tau+h}^{(V)} = \bm{y}_{\tau+h}^{(m-1)}$. The stacked vector is constructed by appending the new draw to the lagged components of its parent and dropping the previous last lag: $\bm{x}_{\tau+h+1}^{(v)} = \big(\,\bm{y}_{\tau+h}^{(v)\prime},\; \bm{y}_{\tau+h-1}^{(a_h^{(v)})\prime},\;\ldots,\;\bm{y}_{\tau+h-p+1}^{(a_h^{(v)})\prime}\,\big)'$.
    \item The final step updates the weights in light of the restrictions given by the measurement equation. In case there are none, we again obtain equal weights. Otherwise, we have $\tilde{w}_{h}^{(v)} \propto \mathcal{N}(\bm{r}_h; \bm{R}_h \bm{\mu}_{\tau+h}^{(v)}, \bm{R}_h\bm{\Sigma}\bm{R}_h' + \bm{\Omega}_h)$.
\end{enumerate}
When the maximum forecast horizon $H$ is reached (i.e., $h = 1,2,\hdots,H$), it remains to generate a fully smoothed trajectory by tracing back through the stored ancestor indices. We draw one final particle index using $v_H\sim\text{Cat}(w_H^{(1)},\hdots,w_H^{(V)})$ and then recursively for each $h = H-1,H-2,\hdots,1$ trace its lineage, setting $v_h = a_{h+1}^{(v_{h + 1})}$. A draw from the distribution $p(\bm{y}_{\tau+1:\tau+H} \cond \mathcal{I}, \mathcal{C}_{1:H},\bm{\Xi})$ is then given by the vector $\bm{y}_{\tau+1:\tau+H}^{(m)}=(\bm{y}_{\tau+1}^{(v_1)\prime},\hdots,\bm{y}_{\tau+H}^{(v_H)\prime})'$, which serves as the reference trajectory in the next sweep of the MCMC algorithm.

To compute GIRFs, we typically require the expectation of the respective scenario and baseline forecast distributions rather than a draw from them. The weights $w_h^{(v)}$ relate to the forward filtering distributions, and do not encode information over the full forecast horizon, so they cannot be used without adjustment. To propagate information backwards, we define smoothing weights, $s_h^{(v)}$, which are initialized as $s_H^{(v)} = w_{H}^{(v)}$. For $h = H-1,\hdots,1,$ we compute $\tilde{s}_{h}^{(v)} = \sum_{j = 1}^{V} s_{h+1}^{(j)} \mathbb{I}(a_{h+1}^{(j)} = v)$ and normalize $s_h^{(v)} = \tilde{s}_{h}^{(v)} / (\sum_{j=1} \tilde{s}_{h}^{(j)})$. We use these smoothed weights to compute the required expectations at each horizon:
\begin{equation*}
    \mathbb{E}(\bm{y}_{t+h} \cond \mathcal{C}_{1:H},\bm{x}_{\tau+1},\mathcal{I},\bm{\Xi}) \approx \sum_{v=1}^V s_h^{(v)} \bm{y}_{\tau+h}^{(v)}.
\end{equation*}

\subsection{Posterior Distributions and Sampling Algorithm}\label{app:posteriorsampling}
We may use the conditional distribution in (\ref{eq:singleeq}) to update the trees equation-by-equation using the backfitting approach designed by \citet{chipman2010bart}; see also \citet{esser2024seemingly}. Here, one may define the vector of partial residuals 
\begin{equation*}
    \tilde{y}_{is,t} = \left(y_{it} - \tilde{\mu}_{it} - \sum_{j\neq s} \ell_{ij}(\bm{x}_t\cond\mathcal{T}_{ij},\bm{\mathrm{m}}_{ij})\right) \sim \mathcal{N}\left(\ell_{is}(\bm{x}_t\cond\mathcal{T}_{is},\bm{\mathrm{m}}_{is}), \varsigma_{it}^2\right),
\end{equation*}
conditioning on the fit of each of the $S-1$ trees except tree $s$ and information in all but the $i$th equation. In full data notation, $\tilde{\bm{y}}_{is} = (\tilde{y}_{is,1},\hdots,\tilde{y}_{is,T})'$, this defines a conditionally Gaussian likelihood, $p(\tilde{\bm{y}}_{is} \cond \mathcal{T}_{is}, \bm{\mathrm{m}}_{is}, \bullet)$, which can be marginalized analytically over the terminal node parameters $\bm{\mathrm{m}}_{is}$ (to keep the dimensionality of the inferential problem fixed). 

Combining this conditional likelihood with the prior on the trees, and a suitable transition density (based on four distinct moves: grow a terminal node, prune a terminal node, change a splitting rule, swap a child/parent node), the trees are sampled using a standard accept/reject Metropolis-Hastings algorithm. These trees (and associated rules) partition the input space and we obtain a distinct set of observations for each terminal node. The posterior then takes the conventional Gaussian form for these parameters.

Updating all trees $s=1,\hdots,S,$ across equations $i = 1,\hdots,n,$ yields an updated fit that can be used to compute the outlier-adjusted residuals $\bm{\epsilon}_t / s_t = \bm{y}_t - \bm{F}(\bm{x}_t)$. The posterior of the constant part of the covariance matrix is then given by:
\begin{equation*}
    \bm{\Sigma} \cond \bullet \sim \mathcal{W}^{-1}\left(s_0 + T, \bm{S}_0 + \sum_{t = 1}^T s_t^{-2}\bm{\epsilon}_t \bm{\epsilon}_t'\right)
\end{equation*}
The hierarchical parameters of the prior on the covariance matrix can be updated using:
\begin{equation*}
    a_i \cond \bullet \sim \mathcal{G}^{-1}\left(\frac{\nu + T}{2}, \frac{1}{A_i^{2}} + \nu \cdot \bm{\Sigma}^{-1}_{[ii]}\right),
\end{equation*}
where $\bm{\Sigma}^{-1}_{[ii]}$ denotes the $i$th diagonal element of $\bm{\Sigma}^{-1}$. The outlier adjustment parameter $s_t$ can be sampled, due to its discrete support, using the probabilities:
\begin{align*}
    \Pr(s_t = 1\cond\bm{y}_t, \bm{F}(\bm{x}_t),\bm{\Sigma},\mathfrak{p}) &\propto \mathcal{N}(\bm{y}_t \cond \bm{F}(\bm{x}_t),\bm{\Sigma}) \cdot (1-\mathfrak{p}),\\
    \Pr(s_t = \mathfrak{s}\cond\bm{y}_t, \bm{F}(\bm{x}_t),\bm{\Sigma},\mathfrak{p}) &\propto \mathcal{N}(\bm{y}_t \cond \bm{F}(\bm{x}_t),\mathfrak{s}^2 \bm{\Sigma}) \cdot \left(\mathfrak{p} / (\overline{\mathfrak{s}} - 1)\right), \quad \text{for } \mathfrak{s} = 2, 3, \hdots, \overline{\mathfrak{s}},
\end{align*}
on a $t$-by-$t$ basis. The posterior distribution of the outlier probability is $\mathfrak{p} \cond \bullet \sim \mathcal{B}(a_\mathfrak{p} + T_{\text{o}},b_\mathfrak{p} + T - T_{\text{o}})$, with the total number of observations classified as outliers denoted by $T_{\text{o}} = \sum_{t = 1}^T \mathbb{I}(s_t \neq 1)$, where $\mathbb{I}(\bullet)$ is an indicator function that yields $1$ if its argument is true and $0$ otherwise.

Note that all model parameters can be updated conditional on any restrictions via data augmentation (by appending the respective draws to the observed data in each MCMC sweep). This involves iteratively alternating between sampling from the (restricted) forecast distribution conditional on the parameters, and vice versa. See also \citet{waggoner1999conditional}, \citet{banbura2015conditional} and \citet{antolin2021structural}.

\subsection{Recursive substitution and structural (G)IRFs}
Note that our model can be written in terms of its one-step ahead predictive distribution, taking the initial conditions $\bm{x}_{\tau+1}$ as given, as $\bm{y}_{\tau+1} = \bm{F}(\bm{x}_{\tau+1}) + \bm{H}^{-1}\bm{u}_{\tau+1}$ with $\bm{u}_{\tau+1}\sim\mathcal{N}(\bm{0}_n, \bm{I}_n)$. In case one desires to integrate over the marginal distributions of the other shocks (when using PGAS, one may simply specify the distribution of the structural shocks explicitly), one may randomly generate two distinct sets of structural shocks \citep[see also][]{kilian2017structural}:
\begin{align*}
    \bm{u}_{j\tau}^{(\tts,m)} &= (u_{1\tau}^{(\tts,m)}, \hdots, u_{j-1,\tau}^{(\tts,m)}, d_0 + d, u_{j+1,\tau}^{(\tts,m)}, \hdots,u_{n\tau}^{(\tts,m)})\\
    \bm{u}_{j\tau}^{(\ttb,m)} &= (u_{1\tau}^{(\ttb,m)}, \hdots, u_{j-1,\tau}^{(\ttb,m)}, d_0, u_{j+1,\tau}^{(\ttb,m)}, \hdots,u_{n\tau}^{(\ttb,m)}).
\end{align*}
When nonlinear impact effects are possible, one may also average over different baseline levels of the shock. These can be used to obtain the GIRF on impact:
\begin{equation*}
    \bm{\delta}_{j\tau,1}^{(d,m)} = \bm{F}(\bm{x}_{\tau+1}) + \bm{H}^{-1}\bm{u}_{j\tau}^{(\tts,m)} - \bm{F}(\bm{x}_{\tau+1}) - \bm{H}^{-1}\bm{u}_{j\tau}^{(\ttb,m)} = \bm{H}^{-1}(\bm{u}_{j\tau}^{(\tts,m)} - \bm{u}_{j\tau}^{(\ttb,m)}).
\end{equation*}
When using the same sequence of random numbers for simulating \tts~and \ttb~(see also the discussion in Appendix \ref{app:empirics}), or when taking expectations in line with our definition of the structural GIRF in (\ref{eq:SGIRF}), we have a closed form expression for the impact response:
\begin{equation*}
    \bm{\delta}_{j\tau,1}^{(d,m)} = \bm{H}^{-1}\left((d_0 + d)\cdot\bm{e}_j' - d_0\cdot\bm{e}_j'\right) = d \cdot\bm{H}^{-1}\bm{e}_j' = d \cdot \bm{\beta}_0^{(m)}, 
\end{equation*}
for iteration $m$ of our sampler. That is, we obtain a scaled version of the $j$th column of the impact matrix $\bm{H}^{-1}$ as in standard VAR models. We thus assume with our additive error specification that the impact of the respective shock is constant over time, and shocks of different signs and sizes are (proportionally) introduced by setting $d$ accordingly. This relates to scaled versions of the average effect for infinitesimally small shocks on impact, $\lim_{d\rightarrow0}\bm{\delta}^{(d)}_{j\tau,1}/d = \bm{\beta}_0^{(m)}$, see also \citet{kolesar2024dynamic}. 

Next, using $\bm{\beta}_0^{(\tts,m)} = \bm{H}^{-1}\bm{u}_{j\tau}^{(\tts,m)}$ and $\bm{\beta}_0^{(\ttb,m)} = \bm{H}^{-1}\bm{u}_{j\tau}^{(\ttb,m)}$ to refer to the impact effects for the scenario and baseline forecasts, we can set $\bm{x}_{\tau+2}^{(\tts,m)} = ((\bm{y}_{\tau+1} + \bm{\beta}_0^{(\tts,m)})',\bm{y}_{\tau}',\hdots,\bm{y}_{\tau-p+2}')'$ and $\bm{x}_{\tau+2}^{(\ttb,m)} = ((\bm{y}_{\tau+1} + \bm{\beta}_0^{(\ttb,m)})',\bm{y}_{\tau}',\hdots,\bm{y}_{\tau-p+2}')'$. As in Section \ref{sec:predictions}, using (\ref{eq:preddisth}), we can iterate forward and obtain $\bm{\delta}_{\tau,2}^{(d,m)} = \bm{F}(\bm{x}_{\tau+2}^{(\tts,m)}) - \bm{F}(\bm{x}_{\tau+2}^{(\ttb,m)})$.\footnote{Note that we may also compute other functions instead of the expectation, e.g., probabilities or quantiles, see also \citet{gallant1993nonlinear}, and more recently, \citet{jordalocal}.} For higher-order responses, iterate forward, draw from the conditional predictive distributions, and reconfigure $\bm{x}_{\tau+h}^{(\tts,m)}$ and $\bm{x}_{\tau+h}^{(\ttb,m)}$ accordingly at each horizon. A draw from the structural GIRF distribution is given by $\bm{\delta}_{\tau,h}^{(d,m)} = \bm{F}(\bm{x}_{\tau+h}^{(\tts,m)}) - \bm{F}(\bm{x}_{\tau+h}^{(\ttb,m)})$. This addresses future reduced form shocks to obtain the respective conditional expectation (implicitly via the recursive simulation of the scenario and baseline paths of the variables). Different to PGAS, this only requires a single realization per MCMC iteration instead of multiple particles.

\inlinehead{Comparison with linear VAR} Under the assumption of Gaussian errors and when $\bm{F}(\bm{y}_t) = \bm{A}\bm{y}_{t}$ is a linear function, where $\bm{A}$ is an $n\times n$ matrix of coefficients, $\bm{y}_{t+h} = \bm{A}^h\bm{y}_t + \sum_{j = 1}^h \bm{A}^{h-j}\bm{\epsilon}_{t+j}$. Conditional on information up to time $t$, we obtain closed form expressions for, e.g., $\mathbb{E}(\bm{y}_{t+h}) = \bm{A}^h \bm{y}_t$ and Var$(\bm{y}_{t+h}) = \sum_{j=1}^h \bm{A}^{h-j}\bm{\Sigma}\bm{A}^{h-j}{'}$. In the nonlinear case, one cannot apply expectations as straightforwardly and compute closed form multi-step ahead expressions, e.g., for the first moment, as $\mathbb{E}(\tilde{\bm{F}}_{(h)}(\bm{y}_t)) \neq \mathbb{E}(\tilde{\bm{F}}_{(h)}(\bm{y}_t,\bm{\epsilon}_{t+1},\bm{\epsilon}_{t+2},\hdots,\bm{\epsilon}_{t+h}))$. The former expectation corresponds to a path which sets the innovations after time $t$ to zero which is typically undesirable, see, e.g., \citet{potter2000nonlinear} for a discussion.

Our approach yields standard IRFs when $\bm{F}(\bullet)$ is linear --- and when all future structural shocks are set equal to $0$ (i.e., the model is iterated in expectations) or when the same randomly generated numbers are used for simulating the scenario and baseline distributions; otherwise they feature uncertainty but are equal in expectation. For instance, when $\bm{F}$ is a linear mapping, we obtain the IRF at horizon $h$ as $d \bm{A}^h \bm{\beta}_0$. That is, it can be obtained by simply projecting the shock impact forward using powers of $\bm{A}$. Since $d$ factors out and the initial conditions cancel, the IRFs are time-invariant and proportional for different shock sizes and signs. In the more general nonlinear context, this is clearly not the case.

\FloatBarrier\section{Data and results}\label{app:empirics}
\subsection{Simulation-based results}
\inlinehead{Conditioning on observables} To assess the performance of our PGAS approach relative to closed form solutions, we provide a comparison with the precision sampler (PS) of \citet{chan2023conditional}. Assuming a linear conditional mean function allows to contrast both algorithms one-to-one (we cannot use the PS for nonlinear implementations, for the reasons described in the paper). We use a homoskedastic linear VAR model with $n = 5$ and $p = 5$ using a conjugate Minnesota prior with hierarchically estimated hyperparameters, as in \citet{giannone2015prior}, to calibrate the parameters used for simulating artificial data. The data used for calibration features output, inflation, employment (\texttt{GDPC1}, \texttt{CPIAUCSL}, \texttt{PAYEMS}; all in annualized log-differences), interest rate (\texttt{FEDFUNDS}) and stock returns (\texttt{SP500}; log differences), reflecting key variables from our applications with real data. 

We use the posterior median estimates of the dynamic coefficients and the covariance matrix, and simulate $T = 1,000$ observations (to limit any effects of the initial conditions). We take the calibrated parameters as given and use the final $p$ of the observations as initial conditions to compute $3,000$ draws from the respective conditional forecast distributions. We impose three distinct restrictions at different horizons, to illustrate how our approach is capable of enforcing the restrictions over the full desired period: (i) the interest rate increases by one unconditional SD at $h = 1$, (ii) inflation is restricted to its unconditional mean for $h = 9,\hdots,12,$ and (iii) output increases by one unconditional SD relative to its initial level at $h = 20$.

\begin{figure}[ht!]
    \includegraphics[width=\textwidth]{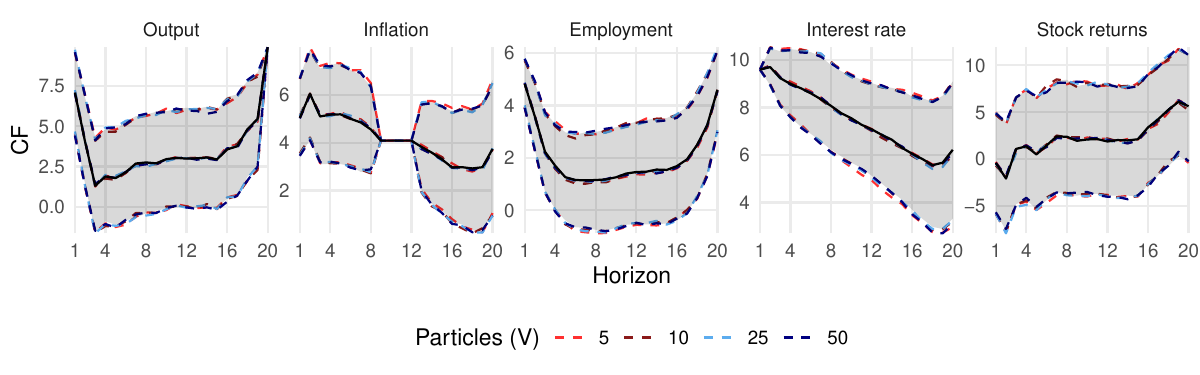}
    \caption{Conditional forecast (CF) with precision sampling (PS) and particle Gibbs with ancestor sampling (PGAS).}\vspace*{-0.25cm}
    \caption*{\footnotesize \textit{Notes}: We obtain 3,000 draws from the distribution of the conditional forecast. The black lines and grey shaded ribbon indicate the median and 68 percent credible set of the predictive distribution obtained from PS. The colored lines indicate draws from PGAS for different numbers of particles $V\in\{5,10,25,50\}$.}
    \label{fig:mcsim}
\end{figure}

The resulting CFs are shown in Figure \ref{fig:mcsim}. The black lines and grey shaded ribbon indicate the median and 68 percent credible set of the predictive distribution obtained from PS. The colored lines indicate draws from PGAS for different numbers of particles for the same quantiles. Differences between implementations are muted, and as little as $5$ particles are sufficient. On a 2020 Macbook Air M1, it takes about $5.5$ seconds with our PS implementation to obtain $1,000$ draws; PGAS runtimes in seconds are 9 (1.6 times as long as PS), 16.6 (3), 41 (7.5), 79.5 (14.5) for particles $V\in\{5,10,25,50\}$.

\begin{figure}[ht!]
    \begin{subfigure}{\textwidth}
    \centering
        \caption{Distinct draws of random numbers (per iteration) for scenario and baseline forecast}
        \includegraphics[width=\textwidth]{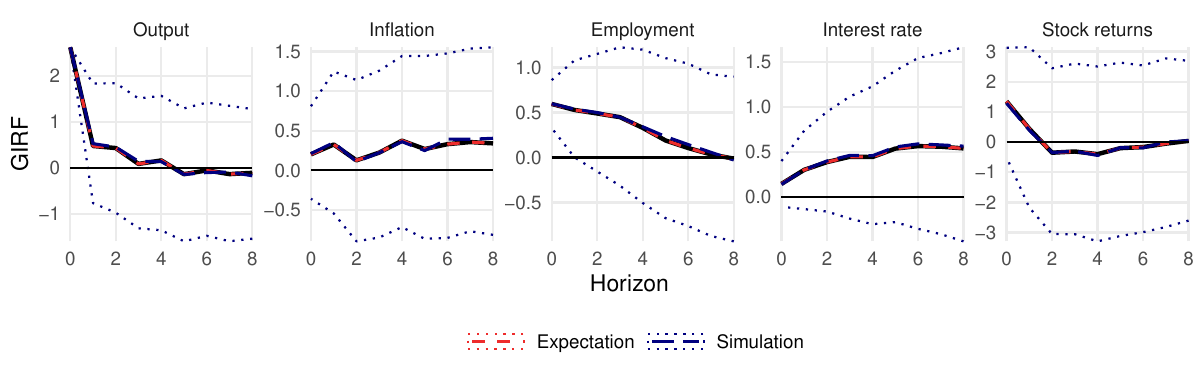}
    \end{subfigure}
    \begin{subfigure}{\textwidth}
    \centering
        \caption{Identical draws of random numbers (per iteration) for scenario and baseline forecast}
        \includegraphics[width=\textwidth]{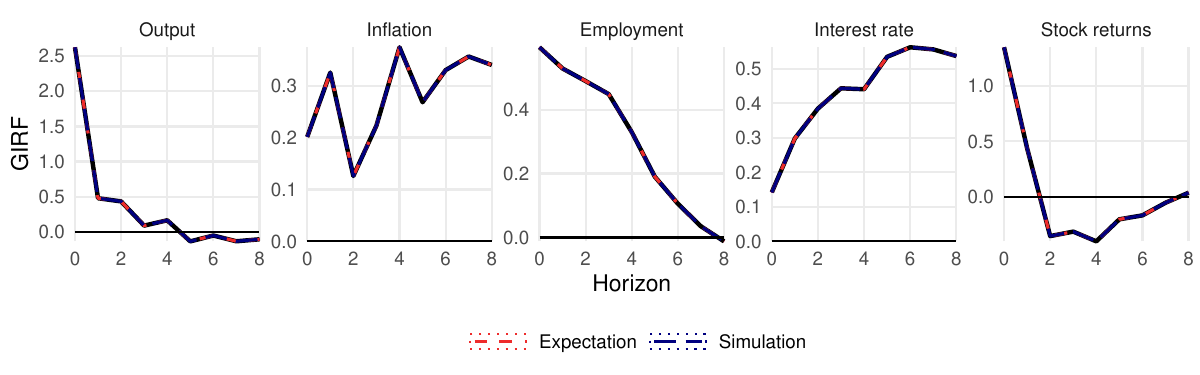}
    \end{subfigure}
    \caption{Generalized impulse response function (IRF) computed with simulation-based methods for a single DGP.}\vspace*{-0.25cm}
    \caption*{\footnotesize \textit{Notes}: When applicable we obtain 3,000 draws from the distribution of the GIRF. The thick black line is the true IRF, colored lines refer to different sampling-based estimation procedures; the thin dotted lines refer to the 68 percent credible set. When using PGAS we rely on $V = 10$ particles.}
    \label{fig:mcsim_irf}
\end{figure}

\inlinehead{Conditioning on shocks: GIRF computation} We use the same DGP as above, and treat the impact matrix $\bm{H}$ as well as the dynamic coefficients as given. This allows to compute the true impulse response function, which is shown in Figure \ref{fig:mcsim_irf} as thick black line. We focus on the response of all variables to the first structural shock $j = 1$. Panel (a) shows the distribution of predictive draws when using different random numbers for the scenario and baseline forecast, while panel (b) uses an identical sequence of random numbers for both forecasts in each iteration. In the latter case, there is no Monte Carlo uncertainty (see also Appendix \ref{app:technical}).

We use PGAS with $V=10$ different particles, subject to distinct choices about the contemporaneous and future shock restrictions. For the variant labeled ``Expectation,'' we set the shock of interest equal to $1$ and all others zero, while we impose these restrictions as approximately binding (with a variance of $10^{-8}$). That is, we iterate in expectations, and obtain exactly the true IRF. For ``Simulation,'' we use $\bm{r}_1^{(u)} = (1,0,\hdots,0)'$ in $\mathcal{C}_1^{(\tts)}$ and $\bm{r}_1^{(u)} = \bm{0}_n$ in $\mathcal{C}_1^{(\ttb)}$, with $\bm{\Omega}_1 = \diag(10^{-8},1,\hdots,1)$ in both cases.

\subsection{Dataset}
Our dataset is roughly patterned after \citet{crump2021large}. For reference, see also the data used in the \href{https://www.federalreserve.gov/supervisionreg/stress-tests-capital-planning.htm}{Dodd-Frank Act stress test}. Variable codes, a brief description and the respective transformations are listed in Table \ref{tab:data}.

\begin{table}[ht]
\caption{Variable codes, descriptions and transformation of our dataset.}\label{tab:data}
\centering
\begin{tabular}{llccc}
  \toprule
\textbf{Code} & \textbf{Description} & $h(x_t)$ & CF & GIRF \\ 
  \midrule
  GDPC1 & Real gross domestic product & 1 & \checkmark & \checkmark \\ 
  PCECC96 & Real personal consumption expenditure & 1 & \checkmark &  \\ 
  PRFIx & Real private fixed investment (residential) & 1 & \checkmark &  \\ 
  GCEC1 & Real Government consumption and investment & 1 & \checkmark &  \\ 
  RDI & Real disposable income & 1 & \checkmark &  \\ 
  INDPRO & Industrial production & 1 & \checkmark &  \\ 
  CPIAUCSL & Headline CPI & 1 & \checkmark & \checkmark \\ 
  CPILFESL & Core CPI & 1 & \checkmark &  \\ 
  PCECTPI & Headline PCE prices & 1 & \checkmark &  \\ 
  PCEPILFE & Core PCE prices & 1 & \checkmark &  \\ 
  HPI & House price index & 1 & \checkmark & \\ 
  HOUST & Housing starts & 1 & \checkmark &  \\ 
  MR & Mortgage rate & 0 & \checkmark &  \\ 
  PAYEMS & Payroll employment & 1 & \checkmark & \checkmark \\ 
  UNRATE & Unemployment rate & 0 & \checkmark &  \\ 
  FEDFUNDS & Federal funds rate & 0 & \checkmark & \checkmark \\ 
  GS1 & 1-year treasury rate & 0 & \checkmark &  \\ 
  GS10 & 10-year treasury rate & 0 & \checkmark & \checkmark \\ 
  EBP & Excess bond premium & 0 & \checkmark & \checkmark \\ 
  NFCI & National financial conditions index & 0 & \checkmark &  \\ 
  OILPRICEx & Real crude oil prices (WTI) & 2 & \checkmark &  \\ 
  SP500 & S\&P 500 & 2 & \checkmark & \checkmark \\ 
  EXUSUKx & US/UK foreign exchange rate & 3 & \checkmark & \checkmark \\ 
  USDEUR & EU/US foreign exchange rate & 3 &  & \checkmark \\ 
  EARGDP & Real gross domestic product (EA) & 1 &  & \checkmark \\ 
  UKRGDP & Real gross domestic product (UK) & 1 &  & \checkmark \\ 
   \bottomrule
\end{tabular}
\caption*{\footnotesize \textit{Notes}: (0) no transformation $h(x_t) = x_t$; (1) annualized log-differences $h(x_t) = 400\cdot\log(x_t/x_{t-1})$; (2) log-differences $h(x_t) = 100\cdot\log(x_t/x_{t-1})$, (3) logarithm $h(x_t) = \log(x_t)$. Check marks indicate inclusion in the information sets for our applications (CF for Sections \ref{sec:dfst} and \ref{sec:nfcitails}, GIRF for Section \ref{sec:finshock}).}
\end{table}

\clearpage\FloatBarrier
\subsection{Empirical results}
\begin{figure}[!ht]
    \centering
    \begin{subfigure}{\textwidth}
    \centering
        \caption{\texttt{BART-hom}}
        \includegraphics[width=0.79\linewidth]{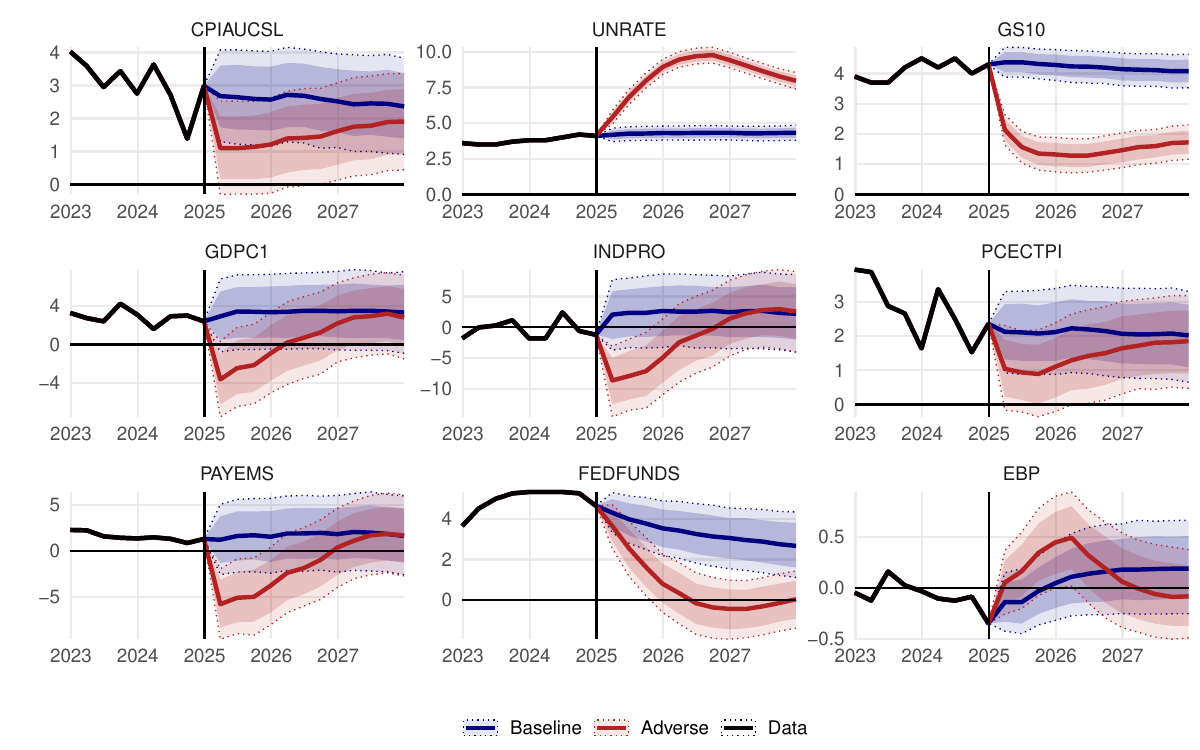}
    \end{subfigure}
    \begin{subfigure}{\textwidth}
    \centering
        \caption{\texttt{BVAR-het}}
        \includegraphics[width=0.79\linewidth]{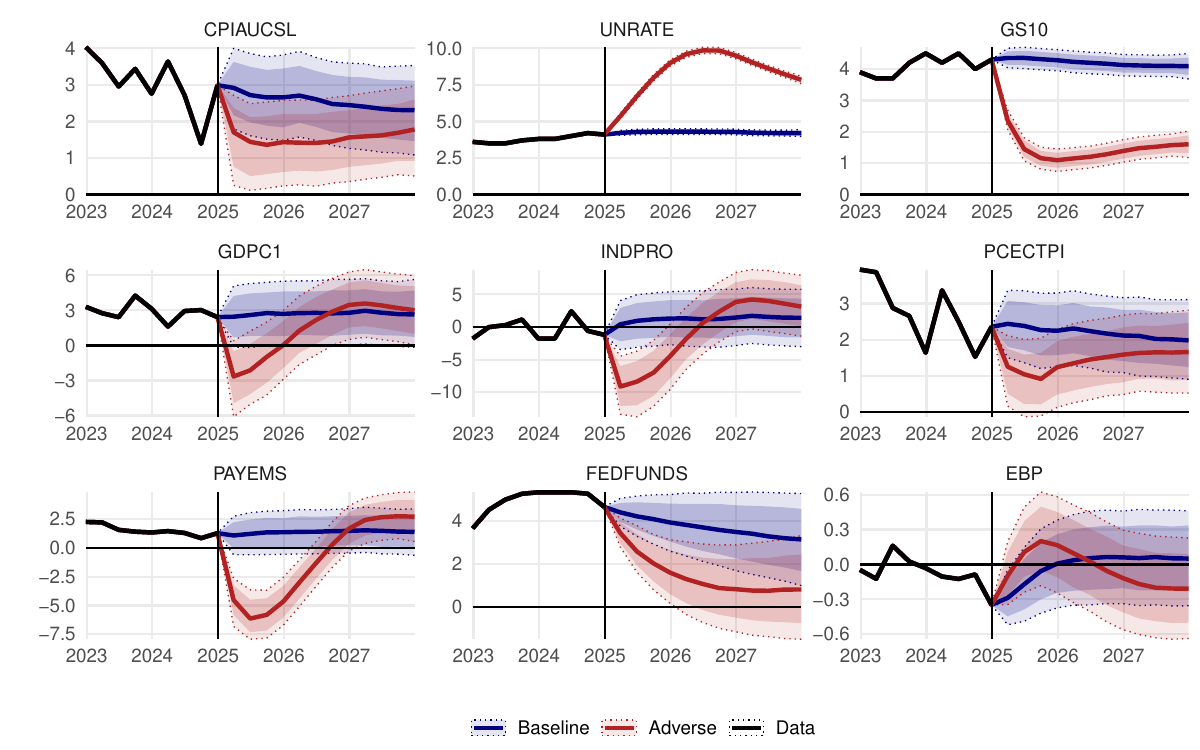}
    \end{subfigure}
    \caption{Conditional forecasts for selected variables.}\vspace*{-0.25cm}
    \caption*{\footnotesize \textit{Notes}: Posterior median alongside 50/68 percent credible sets. Restricted variables: consumer price inflation (\texttt{CPIAUCSL}), unemployment rate (\texttt{UNRATE}), $10$-year government bond yields (\texttt{GS10}); unrestricted variables: Real GDP (\texttt{GDPC1}), industrial production (\texttt{INDPRO}), personal consumption expenditure inflation (\texttt{PCECTPI}), payroll employment (\texttt{PAYEMS}), federal funds rate (\texttt{FEDFUNDS}) and excess bond premium (\texttt{EBP}).}
    \label{fig:cf_dfst_others}
\end{figure}

\begin{figure}[t]
    \centering
    \includegraphics[width=\linewidth]{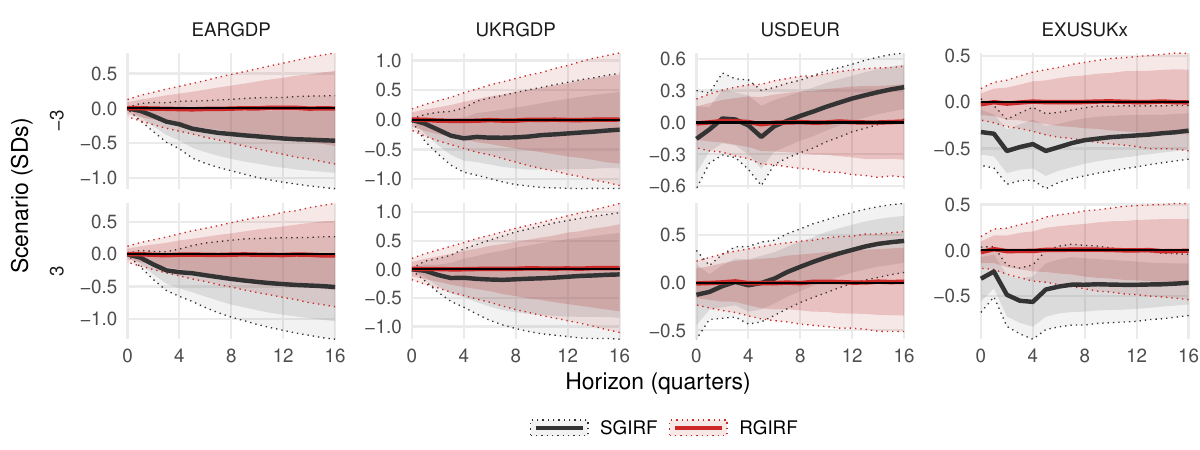}
    \caption{Structural and restricted GIRFs for non-domestic variables.}\vspace*{-0.25cm}
    \caption*{\footnotesize \textit{Notes}: The restricted case assumes that the US financial shock does not spill over to EA and UK real GDP (\texttt{EARGDP}, \texttt{UKRGDP}) and bilateral exchange rates (euro, \texttt{USDEUR}; British pound, \texttt{EXUSUKx}). Posterior medians alongside 50/68 percent credible sets. Cumulated responses for variables in differences and levels for all other variables.}
    \label{fig:girfs_foreign_diff}
\end{figure}

\FloatBarrier
{\setstretch{1.2}\putbib}
\end{bibunit}
\end{appendices}


\begin{thebibliography}{73}
\newcommand{\enquote}[1]{``#1''}
\providecommand{\natexlab}[1]{#1}

\bibitem[{Adams \emph{et~al.}(2021)Adams, Adrian, Boyarchenko, and
  Giannone}]{adams2021forecasting}
\textsc{Adams PA, Adrian T, Boyarchenko N, and Giannone D} (2021),
  \enquote{Forecasting macroeconomic risks,} \emph{International Journal of
  Forecasting} \textbf{37}(3), 1173--1191.

\bibitem[{Adrian \emph{et~al.}(2019)Adrian, Boyarchenko, and
  Giannone}]{adrian2019vulnerable}
\textsc{Adrian T, Boyarchenko N, and Giannone D} (2019), \enquote{Vulnerable
  growth,} \emph{American Economic Review} \textbf{109}(4), 1263--1289.

\bibitem[{Adrian \emph{et~al.}(2021)Adrian, Boyarchenko, and
  Giannone}]{adrian2021multimodality}
---{}---{}--- (2021), \enquote{Multimodality in macrofinancial dynamics,}
  \emph{International Economic Review} \textbf{62}(2), 861--886.

\bibitem[{Adrian \emph{et~al.}(2025)Adrian, Giannone, Luciani, and
  West}]{adrian2025scenario}
\textsc{Adrian T, Giannone D, Luciani M, and West M} (2025), \enquote{Scenario
  Synthesis and Macroeconomic Risk,} \emph{arXiv} \textbf{2505.05193}.

\bibitem[{Alessandri and Mumtaz(2019)}]{alessandri2019financial}
\textsc{Alessandri P, and Mumtaz H} (2019), \enquote{Financial regimes and
  uncertainty shocks,} \emph{Journal of Monetary Economics} \textbf{101},
  31--46.

\bibitem[{Andersson \emph{et~al.}(2010)Andersson, Palmqvist, and
  Waggoner}]{andersson2010density}
\textsc{Andersson MK, Palmqvist S, and Waggoner DF} (2010),
  \enquote{Density-conditional forecasts in dynamic multivariate models,}
  \emph{Sveriges Riksbank Working Paper Series} \textbf{247}.

\bibitem[{Andrieu \emph{et~al.}(2010)Andrieu, Doucet, and
  Holenstein}]{Andrieu2010}
\textsc{Andrieu C, Doucet A, and Holenstein R} (2010), \enquote{{Particle
  Markov chain Monte Carlo methods},} \emph{Journal of the Royal Statistical
  Society Series B} \textbf{72}(3), 269--342.

\bibitem[{Antolin-Diaz \emph{et~al.}(2021)Antolin-Diaz, Petrella, and
  Rubio-Ram{\'\i}rez}]{antolin2021structural}
\textsc{Antolin-Diaz J, Petrella I, and Rubio-Ram{\'\i}rez JF} (2021),
  \enquote{Structural scenario analysis with SVARs,} \emph{Journal of Monetary
  Economics} \textbf{117}, 798--815.

\bibitem[{Arias \emph{et~al.}(2023)Arias, Rubio-Ramirez, and
  Shin}]{arias2023macroeconomic}
\textsc{Arias JE, Rubio-Ramirez JF, and Shin M} (2023), \enquote{Macroeconomic
  forecasting and variable ordering in multivariate stochastic volatility
  models,} \emph{Journal of Econometrics} \textbf{235}(2), 1054--1086.

\bibitem[{Ba{\'n}bura \emph{et~al.}(2015)Ba{\'n}bura, Giannone, and
  Lenza}]{banbura2015conditional}
\textsc{Ba{\'n}bura M, Giannone D, and Lenza M} (2015), \enquote{Conditional
  forecasts and scenario analysis with vector autoregressions for large
  cross-sections,} \emph{International Journal of Forecasting} \textbf{31}(3),
  739--756.

\bibitem[{Barnichon \emph{et~al.}(2022)Barnichon, Matthes, and
  Ziegenbein}]{barnichon2022effects}
\textsc{Barnichon R, Matthes C, and Ziegenbein A} (2022), \enquote{Are the
  effects of financial market disruptions big or small?} \emph{Review of
  Economics and Statistics} \textbf{104}(3), 557--570.

\bibitem[{Baumeister and Benati(2013)}]{baumeister2012unconventional}
\textsc{Baumeister C, and Benati L} (2013), \enquote{Unconventional Monetary
  Policy and the Great Recession: Estimating the Macroeconomic Effects of a
  Spread Compression at the Zero Lower Bound,} \emph{International Journal of
  Central Banking} \textbf{31}.

\bibitem[{Baumeister and Kilian(2014)}]{baumeister2014real}
\textsc{Baumeister C, and Kilian L} (2014), \enquote{Real-time analysis of oil
  price risks using forecast scenarios,} \emph{IMF Economic Review}
  \textbf{62}(1), 119--145.

\bibitem[{Baumeister and Peersman(2013)}]{baumeister2013time}
\textsc{Baumeister C, and Peersman G} (2013), \enquote{Time-varying effects of
  oil supply shocks on the US economy,} \emph{American Economic Journal:
  Macroeconomics} \textbf{5}(4), 1--28.

\bibitem[{Bernanke \emph{et~al.}(1997)Bernanke, Gertler, Watson, Sims, and
  Friedman}]{bernanke1997systematic}
\textsc{Bernanke BS, Gertler M, Watson M, Sims CA, and Friedman BM} (1997),
  \enquote{Systematic monetary policy and the effects of oil price shocks,}
  \emph{Brookings Papers on Economic Activity} \textbf{1997}(1), 91--157.

\bibitem[{Breitenlechner \emph{et~al.}(2024)Breitenlechner, Geiger, and
  Klein}]{breitenlechner2024fiscal}
\textsc{Breitenlechner M, Geiger M, and Klein M} (2024), \enquote{The fiscal
  channel of monetary policy,} \emph{University of Innsbruck WPs in Economics
  and Statistics} \textbf{2024-07}.

\bibitem[{Breitenlechner \emph{et~al.}(2022)Breitenlechner, Georgiadis, and
  Schumann}]{breitenlechner2022goes}
\textsc{Breitenlechner M, Georgiadis G, and Schumann B} (2022), \enquote{What
  goes around comes around: How large are spillbacks from US monetary policy?}
  \emph{Journal of Monetary Economics} \textbf{131}, 45--60.

\bibitem[{Carriero \emph{et~al.}(2024)Carriero, Clark, Marcellino, and
  Mertens}]{carriero2021addressing}
\textsc{Carriero A, Clark TE, Marcellino M, and Mertens E} (2024),
  \enquote{Addressing {COVID-19} outliers in BVARs with stochastic volatility,}
  \emph{Review of Economics and Statistics} 1--15.

\bibitem[{Carriero \emph{et~al.}(2025)Carriero, Clark, Marcellino, and
  Mertens}]{carriero2023shadow}
---{}---{}--- (2025), \enquote{Forecasting with shadow rate VARs,}
  \emph{Quantitative Economics} \textbf{16}(3), 795--822.

\bibitem[{Chan(2023)}]{chan2023comparing}
\textsc{Chan JC} (2023), \enquote{Comparing stochastic volatility
  specifications for large Bayesian VARs,} \emph{Journal of Econometrics}
  \textbf{235}(2), 1419--1446.

\bibitem[{Chan \emph{et~al.}(2024)Chan, Koop, and Yu}]{chan2024large}
\textsc{Chan JC, Koop G, and Yu X} (2024), \enquote{Large order-invariant
  Bayesian VARs with stochastic volatility,} \emph{Journal of Business \&
  Economic Statistics} \textbf{42}(2), 825--837.

\bibitem[{Chan \emph{et~al.}(2025)Chan, Pettenuzzo, Poon, and
  Zhu}]{chan2023conditional}
\textsc{Chan JC, Pettenuzzo D, Poon A, and Zhu D} (2025), \enquote{Conditional
  Forecasts in Large Bayesian VARs with Multiple Equality and Inequality
  Constraints,} \emph{Journal of Economic Dynamics and Control}
  \textbf{105061}.

\bibitem[{Chernis \emph{et~al.}(2025)Chernis, Hauzenberger, Mumtaz, and
  Pfarrhofer}]{chernis2025bayesian}
\textsc{Chernis T, Hauzenberger N, Mumtaz H, and Pfarrhofer M} (2025),
  \enquote{A Bayesian Gaussian Process Dynamic Factor Model,} \emph{arXiv}
  \textbf{2509.04928}.

\bibitem[{Chernis \emph{et~al.}(2024)Chernis, Koop, Tallman, and
  West}]{chernis2024decision}
\textsc{Chernis T, Koop G, Tallman E, and West M} (2024), \enquote{Decision
  synthesis in monetary policy,} \emph{arXiv} \textbf{2406.03321}.

\bibitem[{Chipman \emph{et~al.}(2010)Chipman, George, and
  McCulloch}]{chipman2010bart}
\textsc{Chipman HA, George EI, and McCulloch RE} (2010), \enquote{BART:
  Bayesian additive regression trees,} \emph{The Annals of Applied Statistics}
  \textbf{4}(1), 266--298.

\bibitem[{Cimadomo \emph{et~al.}(2022)Cimadomo, Giannone, Lenza, Monti, and
  Sokol}]{cimadomo2022nowcasting}
\textsc{Cimadomo J, Giannone D, Lenza M, Monti F, and Sokol A} (2022),
  \enquote{Nowcasting with large Bayesian vector autoregressions,}
  \emph{Journal of Econometrics} \textbf{231}(2), 500--519.

\bibitem[{Clark \emph{et~al.}(2025)Clark, Huber, and
  Koop}]{clark2025nonparametric}
\textsc{Clark T, Huber F, and Koop G} (2025), \enquote{A Nonparametric Approach
  to Augmenting a Bayesian VAR with Nonlinear Factors,} \emph{arXiv}
  \textbf{2508.13972}.

\bibitem[{Clark \emph{et~al.}(2023)Clark, Huber, Koop, Marcellino, and
  Pfarrhofer}]{clark2021tail}
\textsc{Clark TE, Huber F, Koop G, Marcellino M, and Pfarrhofer M} (2023),
  \enquote{Tail forecasting with multivariate Bayesian additive regression
  trees,} \emph{International Economic Review} \textbf{64}(3), 979--1022.

\bibitem[{Clark \emph{et~al.}(2024)Clark, Huber, Koop, Marcellino, and
  Pfarrhofer}]{clark2024investigating}
---{}---{}--- (2024), \enquote{Investigating growth-at-risk using a
  multicountry nonparametric quantile factor model,} \emph{Journal of Business
  \& Economic Statistics} \textbf{42}(4), 1302--1317.

\bibitem[{Cong \emph{et~al.}(2017)Cong, Chen, and Zhou}]{cong2017fast}
\textsc{Cong Y, Chen B, and Zhou M} (2017), \enquote{Fast simulation of
  hyperplane-truncated multivariate normal distributions,} \emph{Bayesian
  Analysis} \textbf{12}(4), 1017--1037.

\bibitem[{Crump \emph{et~al.}(2025)Crump, Eusepi, Giannone, Qian, and
  Sbordonea}]{crump2021large}
\textsc{Crump RK, Eusepi S, Giannone D, Qian E, and Sbordonea A} (2025),
  \enquote{A Large Bayesian VAR of the US Economy,} \emph{International Journal
  of Central Banking} \textbf{21}(2), 351--409.

\bibitem[{Doan \emph{et~al.}(1984)Doan, Litterman, and
  Sims}]{doan1984forecasting}
\textsc{Doan T, Litterman R, and Sims C} (1984), \enquote{Forecasting and
  conditional projection using realistic prior distributions,}
  \emph{Econometric Reviews} \textbf{3}(1), 1--100.

\bibitem[{Esser \emph{et~al.}(2024)Esser, Maia, Parnell, Bosmans, van Dongen,
  Klausch, and Murphy}]{esser2024seemingly}
\textsc{Esser J, Maia M, Parnell AC, Bosmans J, van Dongen H, Klausch T, and
  Murphy K} (2024), \enquote{Seemingly unrelated Bayesian additive regression
  trees for cost-effectiveness analyses in healthcare,} \emph{arXiv}
  \textbf{2404.02228}.

\bibitem[{Fischer \emph{et~al.}(2023)Fischer, Hauzenberger, Huber, and
  Pfarrhofer}]{fischer2023general}
\textsc{Fischer MM, Hauzenberger N, Huber F, and Pfarrhofer M} (2023),
  \enquote{General Bayesian time-varying parameter vector autoregressions for
  modeling government bond yields,} \emph{Journal of Applied Econometrics}
  \textbf{38}(1), 69--87.

\bibitem[{Forni \emph{et~al.}(2024)Forni, Gambetti, Maffei-Faccioli, and
  Sala}]{forni2024nonlinear}
\textsc{Forni M, Gambetti L, Maffei-Faccioli N, and Sala L} (2024),
  \enquote{Nonlinear transmission of financial shocks: Some new evidence,}
  \emph{Journal of Money, Credit and Banking} \textbf{56}(1), 5--33.

\bibitem[{Gallant \emph{et~al.}(1993)Gallant, Rossi, and
  Tauchen}]{gallant1993nonlinear}
\textsc{Gallant AR, Rossi PE, and Tauchen G} (1993), \enquote{Nonlinear dynamic
  structures,} \emph{Econometrica} \textbf{61}(4), 871--907.

\bibitem[{Geweke and Amisano(2010)}]{geweke2010comparing}
\textsc{Geweke J, and Amisano G} (2010), \enquote{Comparing and evaluating
  Bayesian predictive distributions of asset returns,} \emph{International
  Journal of Forecasting} \textbf{26}(2), 216--230.

\bibitem[{Gilchrist and Zakraj{\v{s}}ek(2012)}]{gilchrist2012credit}
\textsc{Gilchrist S, and Zakraj{\v{s}}ek E} (2012), \enquote{Credit spreads and
  business cycle fluctuations,} \emph{American Economic Review}
  \textbf{102}(4), 1692--1720.

\bibitem[{Godsill \emph{et~al.}(2004)Godsill, Doucet, and
  West}]{godsill2004monte}
\textsc{Godsill SJ, Doucet A, and West M} (2004), \enquote{Monte Carlo
  smoothing for nonlinear time series,} \emph{Journal of the American
  Statistical Association} \textbf{99}(465), 156--168.

\bibitem[{Gon{\c{c}}alves \emph{et~al.}(2021)Gon{\c{c}}alves, Herrera, Kilian,
  and Pesavento}]{gonccalves2021impulse}
\textsc{Gon{\c{c}}alves S, Herrera AM, Kilian L, and Pesavento E} (2021),
  \enquote{Impulse response analysis for structural dynamic models with
  nonlinear regressors,} \emph{Journal of Econometrics} \textbf{225}(1),
  107--130.

\bibitem[{Gon{\c{c}}alves \emph{et~al.}(2024)Gon{\c{c}}alves, Herrera, Kilian,
  and Pesavento}]{gonccalves2024state}
---{}---{}--- (2024), \enquote{State-dependent local projections,}
  \emph{Journal of Econometrics} \textbf{244}(2), 105702.

\bibitem[{Goulet~Coulombe(2024)}]{goulet2024bag}
\textsc{Goulet~Coulombe P} (2024), \enquote{To bag is to prune,} \emph{Studies
  in Nonlinear Dynamics \& Econometrics} \textbf{forthcoming}.

\bibitem[{Goulet~Coulombe \emph{et~al.}(2022)Goulet~Coulombe, Leroux,
  Stevanovic, and Surprenant}]{goulet2022machine}
\textsc{Goulet~Coulombe P, Leroux M, Stevanovic D, and Surprenant S} (2022),
  \enquote{How is machine learning useful for macroeconomic forecasting?}
  \emph{Journal of Applied Econometrics} \textbf{37}(5), 920--964.

\bibitem[{Gourieroux and Lee(2023)}]{gourieroux2023nonlinear}
\textsc{Gourieroux C, and Lee Q} (2023), \enquote{Nonlinear impulse response
  functions and local projections,} \emph{arXiv} \textbf{2305.18145}.

\bibitem[{Hamilton and Herrera(2004)}]{hamilton2004comment}
\textsc{Hamilton JD, and Herrera AM} (2004), \enquote{Comment: ``Oil shocks and
  aggregate macroeconomic behavior: the role of monetary policy'',}
  \emph{Journal of Money, Credit and Banking} 265--286.

\bibitem[{Hauzenberger \emph{et~al.}(2025{\natexlab{a}})Hauzenberger, Huber,
  Klieber, and Marcellino}]{hauzenberger2024machine}
\textsc{Hauzenberger N, Huber F, Klieber K, and Marcellino M}
  (2025{\natexlab{a}}), \enquote{Machine learning the macroeconomic effects of
  financial shocks,} \emph{Economics Letters} \textbf{250}, 112260.

\bibitem[{Hauzenberger \emph{et~al.}(2024)Hauzenberger, Huber, and
  Koop}]{hauzenberger2024bookchapter}
\textsc{Hauzenberger N, Huber F, and Koop G} (2024), \enquote{Macroeconomic
  forecasting using BVARs,} in \textsc{MP~Clements, and AB~Galv\~{a}o} (eds.)
  \enquote{Handbook of Research Methods and Applications in Macroeconomic
  Forecasting,} chapter~2, 15--42, Cheltenham, UK/Northampton, USA: Edward
  Elgar Publishing.

\bibitem[{Hauzenberger \emph{et~al.}(2025{\natexlab{b}})Hauzenberger, Huber,
  Marcellino, and Petz}]{hauzenberger2024gaussian}
\textsc{Hauzenberger N, Huber F, Marcellino M, and Petz N}
  (2025{\natexlab{b}}), \enquote{Gaussian process vector autoregressions and
  macroeconomic uncertainty,} \emph{Journal of Business \& Economic Statistics}
  \textbf{43}(1), 27--43.

\bibitem[{Huber \emph{et~al.}(2024)Huber, Klieber, Marcellino, Onorante, and
  Pfarrhofer}]{huber2024asymmetries}
\textsc{Huber F, Klieber K, Marcellino MG, Onorante L, and Pfarrhofer M}
  (2024), \enquote{Asymmetries in International Financial Spillovers,}
  \emph{SSRN} \textbf{5054831}.

\bibitem[{Huber \emph{et~al.}(2023)Huber, Koop, Onorante, Pfarrhofer, and
  Schreiner}]{huber2023nowcasting}
\textsc{Huber F, Koop G, Onorante L, Pfarrhofer M, and Schreiner J} (2023),
  \enquote{Nowcasting in a pandemic using non-parametric mixed frequency VARs,}
  \emph{Journal of Econometrics} \textbf{232}(1), 52--69.

\bibitem[{Huber and Rossini(2022)}]{huber2022inference}
\textsc{Huber F, and Rossini L} (2022), \enquote{Inference in Bayesian additive
  vector autoregressive tree models,} \emph{The Annals of Applied Statistics}
  \textbf{16}(1), 104--123.

\bibitem[{Jaroci{\'n}ski(2010)}]{jarocinski2010conditional}
\textsc{Jaroci{\'n}ski M} (2010), \enquote{Conditional forecasts and
  uncertainty about forecast revisions in vector autoregressions,}
  \emph{Economics Letters} \textbf{108}(3), 257--259.

\bibitem[{Jord{\`a}(2005)}]{jorda2005estimation}
\textsc{Jord{\`a} {\`O}} (2005), \enquote{Estimation and inference of impulse
  responses by local projections,} \emph{American Economic Review}
  \textbf{95}(1), 161--182.

\bibitem[{Jord{\`a} and Taylor(2025)}]{jordalocal}
\textsc{Jord{\`a} {\`O}, and Taylor AM} (2025), \enquote{Local projections,}
  \emph{Journal of Economic Literature} \textbf{63}(1), 59--110.

\bibitem[{Kilian and L{\"u}tkepohl(2017)}]{kilian2017structural}
\textsc{Kilian L, and L{\"u}tkepohl H} (2017), \emph{Structural vector
  autoregressive analysis}, Cambridge University Press.

\bibitem[{Koles\'{a}r and Plagborg-M{\o}ller(2025)}]{kolesar2024dynamic}
\textsc{Koles\'{a}r M, and Plagborg-M{\o}ller M} (2025), \enquote{Dynamic
  Causal Effects in a Nonlinear World: the Good, the Bad, and the Ugly,}
  \emph{Journal of Business \& Economic Statistics} \textbf{43}(4), 737--754.

\bibitem[{Koop \emph{et~al.}(1996)Koop, Pesaran, and Potter}]{koop1996impulse}
\textsc{Koop G, Pesaran MH, and Potter SM} (1996), \enquote{Impulse response
  analysis in nonlinear multivariate models,} \emph{Journal of Econometrics}
  \textbf{74}(1), 119--147.

\bibitem[{Leeper and Zha(2003)}]{leeper2003modest}
\textsc{Leeper EM, and Zha T} (2003), \enquote{Modest policy interventions,}
  \emph{Journal of Monetary Economics} \textbf{50}(8), 1673--1700.

\bibitem[{Lima \emph{et~al.}(2025)Lima, Carvalho, Lopes, and
  Herren}]{lima2025minnesota}
\textsc{Lima PA, Carvalho CM, Lopes HF, and Herren A} (2025),
  \enquote{Minnesota BART,} \emph{arXiv} \textbf{2503.13759}.

\bibitem[{Lindsten \emph{et~al.}(2014)Lindsten, Jordan, and
  Sch{{\"o}}n}]{lindsten14a}
\textsc{Lindsten F, Jordan MI, and Sch{{\"o}}n TB} (2014), \enquote{Particle
  Gibbs with Ancestor Sampling,} \emph{Journal of Machine Learning Research}
  \textbf{15}(63), 2145--2184.

\bibitem[{Lopez-Salido and Loria(2024)}]{lopez2024inflation}
\textsc{Lopez-Salido D, and Loria F} (2024), \enquote{Inflation at risk,}
  \emph{Journal of Monetary Economics} \textbf{145}, 103570.

\bibitem[{Lucas(1976)}]{lucas1976}
\textsc{Lucas RE} (1976), \enquote{Econometric policy evaluation: A critique,}
  \emph{Carnegie-Rochester Conference Series on Public Policy} \textbf{1},
  19--46.

\bibitem[{Marcellino and Pfarrhofer(2024)}]{marcellino2024bookchapter}
\textsc{Marcellino M, and Pfarrhofer M} (2024), \enquote{Bayesian nonparametric
  methods for macroeconomic forecasting,} in \textsc{MP~Clements, and
  AB~Galv\~{a}o} (eds.) \enquote{Handbook of Research Methods and Applications
  in Macroeconomic Forecasting,} chapter~5, 90--125, Cheltenham,
  UK/Northampton, USA: Edward Elgar Publishing.

\bibitem[{McKay and Wolf(2023)}]{mckay2023can}
\textsc{McKay A, and Wolf CK} (2023), \enquote{What can time-series regressions
  tell us about policy counterfactuals?} \emph{Econometrica} \textbf{91}(5),
  1695--1725.

\bibitem[{Medeiros \emph{et~al.}(2021)Medeiros, Vasconcelos, Veiga, and
  Zilberman}]{medeiros2021forecasting}
\textsc{Medeiros MC, Vasconcelos GF, Veiga {\'A}, and Zilberman E} (2021),
  \enquote{Forecasting inflation in a data-rich environment: the benefits of
  machine learning methods,} \emph{Journal of Business \& Economic Statistics}
  \textbf{39}(1), 98--119.

\bibitem[{Mumtaz and Piffer(2025)}]{mumtaz2022impulse}
\textsc{Mumtaz H, and Piffer M} (2025), \enquote{Impulse response estimation
  via flexible local projections,} \emph{Journal of Money, Credit and Banking}
  \textbf{forthcoming}.

\bibitem[{Pfarrhofer(2022)}]{pfarrhofer2022modeling}
\textsc{Pfarrhofer M} (2022), \enquote{Modeling tail risks of inflation using
  unobserved component quantile regressions,} \emph{Journal of Economic
  Dynamics and Control} \textbf{143}, 104493.

\bibitem[{Potter(2000)}]{potter2000nonlinear}
\textsc{Potter SM} (2000), \enquote{Nonlinear impulse response functions,}
  \emph{Journal of Economic Dynamics and Control} \textbf{24}(10), 1425--1446.

\bibitem[{Rambachan and Shephard(2021)}]{rambachan2021common}
\textsc{Rambachan A, and Shephard N} (2021), \enquote{When do common time
  series estimands have nonparametric causal meaning?} \emph{Mimeo} .

\bibitem[{Sims and Zha(2006)}]{sims2006does}
\textsc{Sims CA, and Zha T} (2006), \enquote{Does monetary policy generate
  recessions?} \emph{Macroeconomic Dynamics} \textbf{10}(2), 231--272.

\bibitem[{Waggoner and Zha(1999)}]{waggoner1999conditional}
\textsc{Waggoner DF, and Zha T} (1999), \enquote{Conditional forecasts in
  dynamic multivariate models,} \emph{Review of Economics and Statistics}
  \textbf{81}(4), 639--651.

\bibitem[{West(2024)}]{west2024perspectives}
\textsc{West M} (2024), \enquote{Perspectives on constrained forecasting,}
  \emph{Bayesian Analysis} \textbf{19}(4), 1013--1039.

\bibitem[{West and Harrison(1997)}]{west1997bayesian}
\textsc{West M, and Harrison J} (1997), \emph{Bayesian forecasting and dynamic
  models}, Springer.

\end{thebibliography}


\begin{thebibliography}{13}
\newcommand{\enquote}[1]{``#1''}
\providecommand{\natexlab}[1]{#1}

\bibitem[{Antolin-Diaz \emph{et~al.}(2021)Antolin-Diaz, Petrella, and
  Rubio-Ram{\'\i}rez}]{antolin2021structural}
\textsc{Antolin-Diaz J, Petrella I, and Rubio-Ram{\'\i}rez JF} (2021),
  \enquote{Structural scenario analysis with SVARs,} \emph{Journal of Monetary
  Economics} \textbf{117}, 798--815.

\bibitem[{Ba{\'n}bura \emph{et~al.}(2015)Ba{\'n}bura, Giannone, and
  Lenza}]{banbura2015conditional}
\textsc{Ba{\'n}bura M, Giannone D, and Lenza M} (2015), \enquote{Conditional
  forecasts and scenario analysis with vector autoregressions for large
  cross-sections,} \emph{International Journal of Forecasting} \textbf{31}(3),
  739--756.

\bibitem[{Chan \emph{et~al.}(2025)Chan, Pettenuzzo, Poon, and
  Zhu}]{chan2023conditional}
\textsc{Chan JC, Pettenuzzo D, Poon A, and Zhu D} (2025), \enquote{Conditional
  Forecasts in Large Bayesian VARs with Multiple Equality and Inequality
  Constraints,} \emph{Journal of Economic Dynamics and Control}
  \textbf{105061}.

\bibitem[{Chipman \emph{et~al.}(2010)Chipman, George, and
  McCulloch}]{chipman2010bart}
\textsc{Chipman HA, George EI, and McCulloch RE} (2010), \enquote{BART:
  Bayesian additive regression trees,} \emph{The Annals of Applied Statistics}
  \textbf{4}(1), 266--298.

\bibitem[{Crump \emph{et~al.}(2025)Crump, Eusepi, Giannone, Qian, and
  Sbordonea}]{crump2021large}
\textsc{Crump RK, Eusepi S, Giannone D, Qian E, and Sbordonea A} (2025),
  \enquote{A Large Bayesian VAR of the US Economy,} \emph{International Journal
  of Central Banking} \textbf{21}(2), 351--409.

\bibitem[{Esser \emph{et~al.}(2024)Esser, Maia, Parnell, Bosmans, van Dongen,
  Klausch, and Murphy}]{esser2024seemingly}
\textsc{Esser J, Maia M, Parnell AC, Bosmans J, van Dongen H, Klausch T, and
  Murphy K} (2024), \enquote{Seemingly unrelated Bayesian additive regression
  trees for cost-effectiveness analyses in healthcare,} \emph{arXiv}
  \textbf{2404.02228}.

\bibitem[{Gallant \emph{et~al.}(1993)Gallant, Rossi, and
  Tauchen}]{gallant1993nonlinear}
\textsc{Gallant AR, Rossi PE, and Tauchen G} (1993), \enquote{Nonlinear dynamic
  structures,} \emph{Econometrica} \textbf{61}(4), 871--907.

\bibitem[{Giannone \emph{et~al.}(2015)Giannone, Lenza, and
  Primiceri}]{giannone2015prior}
\textsc{Giannone D, Lenza M, and Primiceri GE} (2015), \enquote{Prior selection
  for vector autoregressions,} \emph{Review of Economics and Statistics}
  \textbf{97}(2), 436--451.

\bibitem[{Jord{\`a} and Taylor(2025)}]{jordalocal}
\textsc{Jord{\`a} {\`O}, and Taylor AM} (2025), \enquote{Local projections,}
  \emph{Journal of Economic Literature} \textbf{63}(1), 59--110.

\bibitem[{Kilian and L{\"u}tkepohl(2017)}]{kilian2017structural}
\textsc{Kilian L, and L{\"u}tkepohl H} (2017), \emph{Structural vector
  autoregressive analysis}, Cambridge University Press.

\bibitem[{Koles\'{a}r and Plagborg-M{\o}ller(2025)}]{kolesar2024dynamic}
\textsc{Koles\'{a}r M, and Plagborg-M{\o}ller M} (2025), \enquote{Dynamic
  Causal Effects in a Nonlinear World: the Good, the Bad, and the Ugly,}
  \emph{Journal of Business \& Economic Statistics} \textbf{43}(4), 737--754.

\bibitem[{Potter(2000)}]{potter2000nonlinear}
\textsc{Potter SM} (2000), \enquote{Nonlinear impulse response functions,}
  \emph{Journal of Economic Dynamics and Control} \textbf{24}(10), 1425--1446.

\bibitem[{Waggoner and Zha(1999)}]{waggoner1999conditional}
\textsc{Waggoner DF, and Zha T} (1999), \enquote{Conditional forecasts in
  dynamic multivariate models,} \emph{Review of Economics and Statistics}
  \textbf{81}(4), 639--651.

\end{thebibliography}
\end{document}